\documentclass[journal, draftcls, onecolumn, 10pt]{IEEEtran}

\usepackage{jdhmacros}
\usepackage[colorlinks]{hyperref}
\usepackage{cleveref}

\newcommand{\Amax}{\mathrm{A_{max}}}
\newcommand{\Xmax}{\mathrm{X_{max}}}
\newcommand{\Xmin}{\mathrm{X_{min}}}
\newtheorem{theorem}{Theorem}[section]
\newtheorem{corollary}{Corollary}[theorem]

\renewcommand\qedsymbol{$\blacksquare$}

\begin{document}

%%%%%%%%% TITLE
\title{\huge Minimax Lower Bounds for Noisy Matrix Completion\\
Under Sparse Factor Models}

\author{Abhinav V. Sambasivan, \emph{Student Member, IEEE} and Jarvis D. Haupt, \emph{Senior Member, IEEE}\thanks{Submitted September 28, 2015.  Revised February 14, 2017 and October 26, 2017. The authors are with the Department of Electrical and Computer Engineering at the University of Minnesota -- Twin Cities. Tel/fax: (612) 625-3300 \ /\ (612) 625-4583. Emails: {\tt \{samba014, jdhaupt\}@umn.edu}. The authors graciously acknowledge support from the DARPA Young Faculty Award, Grant No. N66001-14-1-4047.}}

\maketitle

\begin{abstract}
This paper examines fundamental error characteristics for a general class of matrix completion problems, where the matrix of interest is a product of two a priori unknown matrices, one of which is sparse, and the observations are noisy. 
Our main contributions come in the form of minimax lower bounds for the expected per-element squared error for this problem under several common noise models. Specifically, we analyze scenarios where the corruptions are characterized by additive Gaussian noise or additive heavier-tailed (Laplace) noise, Poisson-distributed observations, and highly-quantized (e.g., one-bit) observations, as instances of our general result.  Our results establish that the error bounds derived in (Soni et al., 2016) for \emph{complexity-regularized maximum likelihood} estimators achieve, up to multiplicative constants and logarithmic factors, the minimax error rates in each of these noise scenarios, provided that the nominal number of observations is large enough, and the sparse factor has (on an average) at least one non-zero per column.
%number of measurements exceeds the ``degrees of freedom'' in the factor model, and the 
%Such models, henceforth referred to as ``sparse factor models", are natural extensions to the low-rank models which have been extensively studied in the realm of matrix completion. The sparse factor models have been widely used in subspace clustering, dictionary learning, and sparse modelling applications. 
\end{abstract}
\vspace{-0.0in}


%%%%%%%%% BODY TEXT
%-------------------------------------------------------------------------
\section{Introduction}\label{sec:intro}
The matrix completion problem involves imputing the missing values of a matrix from an incomplete, and possibly noisy sampling of its entries. In general, without making any assumption about the entries of the matrix, the matrix completion problem is ill-posed and it is impossible to recover the matrix uniquely. However, if the matrix to be recovered has some intrinsic structure (e.g., low rank structure), it is possible to design algorithms that exactly estimate the missing entries. %with high probability. A common approach has been to attribute a \emph{low-dimensional} structure to the underlying matrix and design an estimator which solves a convex optimization problem to compute the missing elements.
%In recent years, there has been considerable research activity in analyzing inference methods for high-dimensional matrices, as such tasks arise in a number of contemporary applications. Collaborative filtering (as in the well known \emph{Netflix-Prize} Challenge \cite{bell2007lessons}) is one such application where the matrix to be estimated consists of an array of users' ratings (or preferences) to a given collection of items (movies and shows). In this scenario, accurately predicting the user ratings for unknown items is a crucial step towards building a recommender system which makes suggestions to users based on their likely ratings to certain items. 
%In this model, each entry of the matrix $\Xb\in\RR^{n_1\times n_2}$, to be estimated is assumed to depend on a small number (say $r\ll n_1,n_2$) of features and is hence expressed as an inner product between two vectors of length $r$.
Indeed, the performance low-rank matrix completion methods have been extensively studied in noiseless settings \cite{candes2009exact, candes2010power, keshavan2010matrix, recht2011simpler, gross2011recovering}, in noisy settings where the observations are affected by additive noise \cite{keshavan2009matrix, candes2010matrix, koltchinskii2011nuclear, rohde2011estimation, cai2013matrix, klopp2014noisy, lafond2015low}, and in settings where the observations are non-linear (e.g., highly-quantized or Poisson distributed observation) functions of the underlying matrix entry (see, \cite{davenport20141, plan2014high, soni2014estimation}). Recent works which explore robust recovery of low-rank matrices under malicious sparse corruptions include \cite{candes2011robust, hsu2011robust, xu2012robust, chen2013low}. 

A notable advantage of using low-rank models is that the estimation strategies involved in completing such matrices can be cast into efficient convex methods which are well-understood and suitable to analyses. The fundamental estimation error characteristics for more general completion problems, for example, those employing general bilinear factor models, have not (to our knowledge) been fully characterized.  In this work we provide several new results in this direction.  Our focus here is on matrix completion problems under \emph{sparse factor model} assumptions, where the matrix to be estimated is well-approximated by a product of two matrices, one of which is sparse.  Such models have been motivated by a variety of applications in dictionary learning, subspace clustering, image demosaicing, and various machine learning problems (see, e.g. the discussion in \cite{soni2014noisy}).   Here, we investigate fundamental lower bounds on the achievable estimation error for these problems in several specific noise scenarios -- additive Gaussian noise, additive heavier-tailed (Laplace) noise, Poisson-distributed observations, and highly-quantized (e.g., one-bit) observations.  Our analyses compliment the upper bounds provided recently in \cite{soni2014noisy} for complexity-penalized maximum likelihood estimation methods, and establish that the error rates obtained in \cite{soni2014noisy} are nearly minimax optimal (as long as the nominal number of measurements is large enough). % For collaborative filtering applications, sparse factor models serve as natural extensions to the low-rank models, where it is reasonable to assume that a user's rating depends only on a small number of features. Sparse factor models capture such dependencies on the latent dictionary factors much better than a low-rank model.

%\subsection{Our Contribution}\label{subsec:contribution}
%%We establish minimax lower bounds on matrix completion problems under sparse factor models. 
%Our main contributions comprise fundamental bounds on the completion of matrices from noisy random measurements.  We examine several scenarios, corresponding to different noise or corruption models.  Specifically, we consider settings characterized by additive Gaussian noise, additive heavier-tailed (Laplace) noise, Poisson-distributed observations, and highly-quantized (e.g., one-bit) observations. Using the results established here, we prove that the \emph{complexity-regularized maximum likelihood} estimator discussed in \cite{soni2014noisy} is minimax optimal up to a logarithmic factor when the sparse factor obeys the linear sparsity property. We also draw comparisons with existing results from the low-rank matrix completion literature, to illustrate how the sparse factor models examined here may be viewed as a more generic set of models. It is observed that the bounds obtained here are lower than their low-rank counterparts which elucidates the potential benefit of leveraging additional structure, here in the form of sparsity of the factors comprising the bilinear model.

%\subsection{Connections with Existing Works}\label{subsec:existing_work}

%\subsection{Outline}
The remainder of the paper is organized as follows. We begin with a brief overview of the various preliminaries and notations in Section \ref{subsec:prelim} followed by a formal definition of the matrix completion problem considered here in Section \ref{sec:prob}. Our main results are stated in Section \ref{sec:result}; there, we establish minimax lower bounds for the recovery of a matrix $\Xb^*$ that admits a sparse factorization under a general class of noise models. We also briefly discuss the implications of these bounds for different instances noise distributions and compare them with existing works. In Section \ref{sec:concl} we conclude with a concise discussion of  possible extensions and potential future directions. The proofs of our main results are provided in the Appendix.
\subsection{Notations and Preliminaries}\label{subsec:prelim}
We provide a brief summary of the notations used here and revisit a few key concepts before delving into our main results. 

We let $a\vee b=\max\{a,b\}$ and $a\wedge b=\min\{a,b\}$. For any $n\in\NN$, $[n]$ denotes the set of integers $\{1,\dots,n\}$. For a matrix $\Mb$, we use the following notation: $\norm{\Mb}_{0}$ denotes the number of non-zero elements in $\Mb$, $\norm{\Mb}_{\infty}=\max_{i,j}|\Mb_{i,j}|$ denotes the entry-wise maximum (absolute) entry of $\Mb$, and $\norm{\Mb}_{F}=\sqrt{\sum_{i,j}\Mb^2_{i,j}}$ denotes the Frobenius norm. We use the standard asymptotic computational complexity ($\cO,\Omega,\Theta$) notations to suppress leading constants in our results for, clarity of exposition.
%\begin{itemize}
%  \item Big O notation used to denote asymptotic upper bounds. We say that $g(x)=\cO(f(x))$ if for any $x_0,\ \exists$ a constant $c>0$ such that $g(x)\leq c\,f(x),\ \forall x>x_0$
%  \item Big Omega notation used to denote asymptotic lower bounds. We say that $g(x)=\Omega(f(x))$ if for any $x_0,\ \exists$ a constant $c>0$ such that $g(x)\geq c\,f(x),\ \forall x>x_0$
%  \item Big Theta notation used to denote asymptotically tight bounds. We say that $g(x)=\Theta(f(x))$ if for any $x_0,\ \exists$ constants $c_1,c_2>0$ such that $c_1\,f(x)\leq g(x)\leq c_2\,f(x),\ \forall x>x_0$
%\end{itemize}

We also briefly recall an important information-theoretic quantity, the Kullback-Leibler divergence (or KL divergence). When $x(z)$ and $y(z)$ denote the pdfs (or pmfs) of real scalar random variables, the KL divergence of $y\text{ from }x$ is denoted by $K(\PP_x,\PP_y)$ and given by
\begin{eqnarray*}
  K(\PP_x,\PP_y)&=& \EE_{Z\sim x(Z)}\left[\log\frac{x(Z)}{y(Z)}\right] \\
  &\triangleq& \EE_{x}\left[\log\frac{x(Z)}{y(Z)}\right],
\end{eqnarray*}
provided $x(z)=0$ whenever $y(z)=0$, and $\infty$ otherwise.  The logarithm is taken be the natural log. %By definition $K(\PP_x,\PP_y)$ is finite only if the support of $x$ is contained in the support of $y$. 
It is worth noting that the KL divergence is not necessarily commutative and, $K(\PP_x,\PP_y)\geq 0$ with $K(\PP_x,\PP_y)=0$ when $x(Z)=y(Z)$. In a sense, the KL divergence quantifies how ``far" apart two distributions are.
\section{Problem Statement}\label{sec:prob}
\subsection{Observation Model}\label{subsec:obs_model}
We consider the problem of estimating the entries of an unknown matrix $\Xb^*\in\RR^{n_1\times n_2}$ that admits a factorization of the form,
\begin{equation}\label{eqn:factn_model}
\Xb^*=\Db^*\Ab^*,
\end{equation}
where for some integer $r\in[n_1 \wedge n_2]$, $\Db^*\in\RR^{n_1\times r}$ and $\Ab^*\in\RR^{r\times n_2}$ are \emph{a priori} unknown factors. Additionally, our focus in this paper will be restricted to the cases where the matrix $\Ab^*$ is $k$-sparse (having no more than $k\leq rn_2$ nonzero elements). We further note here that if $k<r$, then the matrix $\Ab^*$ will necessarily have zero rows which can be removed without affecting the product $\Xb^*$. Hence without loss of generality, we assume that $k\geq r$. In addition to this, we assume that the elements of $\Db^*$ and $\Ab^*$ are bounded, so that
\begin{equation}\label{eqn:bdd_assmptn}
\norm{\Db^*}_{\infty}\leq1 \text{ and } \norm{\Ab^*}_{\infty}\leq \Amax
\end{equation}
for some constant $\Amax > 0$. A direct implication of (\ref{eqn:bdd_assmptn}) is that the elements of $\Xb^*$ are also bounded ($\norm{\Xb^*}_{\infty}\leq \Xmax \leq r\Amax$). However in most applications of interest $\Xmax$ need not be as large as $r\Amax$ (for example, in the case of recommender systems, the entries of $\Xb^*$ are bounded above by constants, i.e. $\Xmax=\cO(1)$). Hence we further assume here that $\Xmax = \Theta(\Amax) = \cO(1)$. While bounds on the amplitudes of the elements of the matrix to be estimated often arise naturally in practice, the assumption that the entries of the factors are bounded fixes some of the scaling ambiguities inherent to the bilinear model.

Instead of observing all the elements of the matrix $\Xb^*$ directly, we assume here that we make noisy observations of $\Xb^*$ at a known \emph{subset} of locations. In what follows, we will model the observations $Y_{i,j}$ as i.i.d draws from a probability distribution (or mass) function parametrized by the true underlying matrix entry $\Xb^*_{i,j}$. We denote by $\cS\subseteq [n_1]\times [n_2]$ the set of locations at which observations are collected, and assume that these points are sampled randomly with $\EE[|\cS|]=m$ (which denotes the nominal number of measurements) for some integer $m$ satisfying $1\leq m\leq n_1n_2$. Specifically, for $\gamma_0=m/(n_1n_2)$, we suppose $\cS$ is generated according to an independent Bernoulli$(\gamma_0)$ model, so that each $(i,j)\in[n_1]\times [n_2]$ is included in $\cS$ independently with probability $\gamma_0$. Thus, given $\cS$, we model the collection of $|\cS|$ measurements of $\Xb^*$ in terms of the collection $\{Y_{i,j}\}_{(i,j)\in \cS}\triangleq\Yb_{\cS}$ of conditionally (on $\cS$) independent random quantities. The joint pdf (or pmf) of the observations can be formally written as
\begin{equation}\label{eqn:pdf_model}
 p_{\Xb^*_{\cS}}(\Yb_{\cS})\triangleq\prod_{(i,j)\in\cS}p_{\Xb^*_{i,j}}(Y_{i,j}) \triangleq \PP_{\Xb^*} \ ,
\end{equation}
where $p_{\Xb^*_{i,j}}(Y_{i,j})$ denotes the corresponding scalar pdf (or pmf), and we use the shorthand $\Xb^*_{\cS}$ to denote the collection of elements of $\Xb^*$ indexed by $(i,j)\in\cS$. Given $\cS$ and the corresponding noisy observations $\Yb_{\cS}$ of $\Xb^*$ distributed according to (\ref{eqn:pdf_model}), our matrix completion problem aims at estimating $\Xb^*$ under the assumption that it admits a sparse factorization as in (\ref{eqn:factn_model}).
\subsection{The minimax risk}
In this paper, we examine the fundamental limits of estimating the elements of a matrix that follows the model (\ref{eqn:factn_model}) and observations as described above, using any possible estimator (irrespective of its computational tractability). The accuracy of an estimator $\hat{\Xb}$ in estimating the entries of the true matrix $\Xb^*$ can be measured in terms of its risk $\cR_{\hat{\Xb}}$ which we define to be the normalized (per-element) Frobenius error,
\begin{equation}\label{eqn:risk}
\cR_{\hat{\Xb}}\triangleq\cfrac{\EE_{\Yb_{\cS}}\left[\norm{\hat{\Xb}-\Xb^*}_{F}^2\right]}{n_1n_2}.
\end{equation}
Here, our notation is meant to denote that the expectation is taken with respect to all of the random quantities (i.e., the joint distribution of $\cS$ and $\Yb_{\cS}$).

Let us now consider a class of matrices parametrized by the inner dimension $r$, sparsity factor $k$ and upper bound $\Amax$ on the amplitude of elements of $\Ab$, where each element in the class obeys the factor model (\ref{eqn:factn_model}) and the assumptions in (\ref{eqn:bdd_assmptn}). Formally, we set
\begin{equation}\label{def:model_cls}
  \cX(r,k,\Amax)\triangleq\{\Xb=\Db\Ab\in\RR^{n_1\times n_2}: \Db\in\RR^{n_1\times r},\  \norm{\Db}_{\infty}\leq1\text{ and } \Ab\in\RR^{r\times n_2},\  \norm{\Ab}_{0}\leq k,\ \norm{\Ab}_{\infty}\leq \Amax \}.
\end{equation}
The worst-case performance of an estimator $\hat{\Xb}$ over the class $\cX(r,k,\Amax)$, under the Frobenius error metric defined in (\ref{eqn:risk}), is given by its maximum risk,
\begin{equation*}
  \tilde{\cR}_{\hat{\Xb}}\triangleq\sup\limits_{\Xb^*\in\cX(r,k,\Amax)} \cR_{\hat{\Xb}}.
\end{equation*}
The estimator having the smallest maximum risk among all possible estimators and is said to achieve the minimax risk, which is a characteristic of the estimation problem itself. For the problem of matrix completion under the sparse factor model described in Section \ref{subsec:obs_model}, the minimax risk is expressed as
\begin{eqnarray}\label{eqn:mm_risk}
  \nonumber \cR^*_{\cX(r,k,\Amax)}&\triangleq& \inf\limits_{\hat{\Xb}} \; \tilde{\cR}_{\hat{\Xb}} \\
  \nonumber &=&\inf\limits_{\hat{\Xb}}\sup\limits_{\Xb^*\in\cX(r,k,\Amax)} \cR_{\hat{\Xb}}\\
  &=&\inf\limits_{\hat{\Xb}}\sup\limits_{\Xb^*\in\cX(r,k,\Amax)} \cfrac{\EE_{\Yb_{\cS}}\left[\norm{\hat{\Xb}-\Xb^*}_{F}^2\right]}{n_1n_2} .
\end{eqnarray}
As we see, the minimax risk depends on the choice of the model class parameters $r$, $k$ and $\Amax$. It is worth noting that inherent in the formulation of the minimax risk are the noise model and the nominal number of observations ($m=\EE[|\cS|]$) made. For the sake of brevity, we shall not make all such dependencies explicit. 

In general it is complicated to obtain closed form solutions for (\ref{eqn:mm_risk}). Here, we will adopt a common approach employed for such problems, and seek to obtain lower bounds on the minimax risk, $\cR^*_{\cX(r,k,\Amax)}$ using tools from \cite{tsybakov2008introduction}. Our analytical approach is inspired also by the approach in \cite{klopp2014robust}, which considered the problem of estimating low-rank matrices corrupted by sparse outliers.  
%-----------------------------------------------------------------------------------
%-----------------------------------------------------------------------------------

\section{Main Results and Implications}\label{sec:result}
In this section we establish lower bounds on the minimax risk for the problem settings defined in Section \ref{sec:prob} where the KL divergences of the associated noise distributions exhibit a certain property (a quadratic upper bound in terms of the underlying parameters of interest; we elaborate on this later). We consider four different noise models; additive Gaussian noise, additive Laplace noise, Poisson noise, and quantized (one-bit) observations, as instances of our general result. The proof of our main result presented in Theorem \ref{thm:general} appears in Appendix \ref{proof:thm:general}.

%\begin{itemize}
%  \item $k<n_2\,$: There is (on an average), less than one non-zero element per column of $\Ab^*$
%  \item $k\geq n_2\,$: There is (on an average), at least one non-zero element per column of $\Ab^*$
%\end{itemize}

\begin{theorem}\label{thm:general}
	Suppose the scalar pdf (or pmf) of the noise distribution satisfies, for all $x,y$ in the domain of the parameter space, 
	\begin{equation}\label{eqn:quad_KLdiv_bound}
		K(\PP_x,\PP_y) \leq \frac{1}{2\mu_D^2} (x-y)^2,
	\end{equation}
	for some constant $\mu_D$ which depends on the distribution.
	For observations made as i.i.d draws $\Yb_{\cS}\sim\PP_{\Xb^*}$, there exist absolute constants $C,\gamma>0$ such that for all $n_1, n_2\geq2$, $r\in[n_1 \wedge n_2]$, and $r\leq k\leq n_1n_2/2$, the minimax risk for sparse factor matrix completion over the model class $\cX(r,k,\Amax)$ obeys
	\begin{equation}\label{eqn:thm:general}
	\cR^*_{\cX(r,k,\Amax)}\geq C\cdot \min\Bigg\{\Delta(k,n_2)\mathrm{A^2_{max}},\gamma^2\:\mu_D^2\bigg(\frac{n_1r+k}{m}\bigg)\!\Bigg\},
	\end{equation}
	where 
	\begin{equation}\label{eqn:Delta}
		\Delta(k,n_2) \triangleq \min\left\{1,(k/n_2) \right\}.
%	\Delta(k,n_2)=\left\{
%	\begin{array}{cc}
%	\left(k/n_2\right) & \mbox{for }  k<n_2 \\
%	1 & \mbox{for } k\geq n_2
%	\end{array}
%	\right. \ .
	\end{equation}
\end{theorem}

 Let us now analyze the result of this theorem more closely and see how the estimation risk varies as a function of the number of measurements obtained, as well as the dimension and sparsity parameters sof the matrix to be estimated. We can look at the minimax risk in equation (\ref{eqn:thm:general}) in two different scenarios w.r.t the sampling regime:
\begin{itemize}
	\item {\bf Large sample regime} or when $m\succeq(n_1r\vee k)$, (where we use the notation $\succeq$ to suppress dependencies on constants). In this case we can rewrite (\ref{eqn:thm:general}) and lower bound the minimax risk as%\footnote{for this we need to follow the line of arguments shown in Appendix \ref{proof:thm:general} with $a_0 \triangleq \gamma_d\left(\mu_D\wedge\Amax\right) \left(\frac{k}{m}\right)^{1/2}$, and $d_0 \triangleq \gamma_d\left(1\wedge\frac{\mu_D}{\Amax}\right) \left(\frac{n_1r}{m}\right)^{1/2}$} 
	\begin{equation}\label{eqn:gen_result:lsr}
		\cR^*_{\cX(r,k,\Amax)}= \Omega\left((\mu_D\wedge\Amax)^2 \bigg(\frac{n_1r+k}{m}\bigg)\Delta(k,n_2)\right).
	\end{equation}
	
	Here the quantities $n_1r$ and $k$ (which give the maximum number of non-zeros in $\Db^*$ and $\Ab^*$ respectively) can be viewed as the number of \textit{degrees of freedom} contributed by each of the factors in the matrix to be estimated. The term $\frac{n_1r}{m}\cdot\Delta(k,n_2)$ can be interpreted as the error associated with the non-sparse factor which follows the parametric rate $(n_1r/m)$ when $k\geq n_2$, i.e. $\Ab^*$ (on an average) has more than one non-zero element per column. Qualitatively, this implies that all the degrees of freedom offered by $\Db^*$ manifest in the estimation of the overall matrix $\Xb^*$ provided there are enough non-zero elements (at least one non-zero per column) in $\Ab^*$. If there are (on an average) less than one non-zero element per column in the sparse factor, a few rows of $\Db^*$ vanish due to the presence of zero columns in $\Ab^*$ and hence all the degrees of freedom in $\Db^*$ are not carried over to $\Xb^*$ (resulting in zero columns in $\Xb^*$). This makes the overall problem easier and reduces the minimax risk (associated with $\Db^*$) by a factor of $(k/n_2)$. Similarly, $\frac{k}{m}\cdot\Delta(k,n_2)$ is the error term associated with the sparse factor $\Ab^*$, and it follows the parametric rate of $(k/m)$ in the large sample regime provided $k\geq n_2$.
	\item {\bf Small sample regime} or when $m\preceq(n_1r\vee k)$. In this case the minimax risk in (\ref{eqn:thm:general}) becomes
	\begin{equation}\label{eqn:gen_result:ssr}
	\cR^*_{\cX(r,k,\Amax)}= \Omega\left(\Delta(k,n_2)\:\mathrm{A^2_{max}}\right).
	\end{equation}  
	Equation (\ref{eqn:gen_result:ssr}) implies that the minimax risk in estimating the unknown matrix doesn't become arbitrarily large when the nominal number of observations is much smaller than the number of \emph{degrees of freedom} in the factors (or when $m\ll n_1r+k$), but is instead lower bounded by the squared-amplitude ($\mathrm{A^2_{max}}$) of the sparse factor (provided there are sufficient non-zeros in $\Ab^*$ for all the \emph{degrees of freedom} to manifest). This is a direct consequence of our assumption that the entries of the factors are bounded.
\end{itemize}

The virtue of expressing the lower bounds for the minimax risk as in (\ref{eqn:thm:general}) is that we don't make any assumptions on the nominal number of samples collected and is hence a valid bound over all sampling regimes. However, in the discussions that follow, we shall often consider the large sample regime and appeal to the lower bounds of the form (\ref{eqn:gen_result:lsr}).  
In the following sections, we consider different noise models which satisfy the KL-divergence criterion (\ref{eqn:quad_KLdiv_bound}) and present lower bounds for each specific instance as corollaries of our general result presented in Theorem \ref{thm:general}.  

\subsection{Additive Gaussian Noise}\label{subsec:AGN}
Let us consider a setting where the observations are corrupted by i.i.d zero-mean additive Gaussian noise with known variance. We have the following result; its proof appears in Appendix \ref{proof:cor:AGN}

\begin{corollary}[\bf{Lower bound for Gaussian Noise}]\label{cor:AGN}
  Suppose $Y_{i,j}=X^*_{i,j}+\xi_{i,j}$, where $\xi_{i,j}$ are i.i.d Gaussian $\cN(0,\sigma^2)$, $\sigma>0$, $\forall (i,j)\in\cS$. There exist absolute constants $C,\gamma>0$ such that for all $n_1, n_2\geq2$, $r\in[n_1 \wedge n_2]$, and $r\leq k\leq n_1n_2/2$, the minimax risk for sparse factor matrix completion over the model class $\cX(r,k,\Amax)$ obeys
  \begin{equation}\label{eqn:cor:AGN}
	\cR^*_{\cX(r,k,\Amax)}\geq C\cdot \min\Bigg\{\Delta(k,n_2)\mathrm{A^2_{max}},\gamma^2\:\sigma^2\bigg(\frac{n_1r+k}{m}\bigg)\!\Bigg\}.
  \end{equation}
\end{corollary}

\begin{remark}\label{remark:iid_AGN}
  If instead of i.i.d Gaussian noise, we have that $\xi_{i,j}$ are just independent zero-mean additive Gaussian random variables with variances $\sigma_{i,j}^2\geq \sigma_{\mathrm{min}}^2\ \forall (i,j)\in\cS$, the result in (\ref{eqn:cor:AGN}) is still valid with the $\sigma$ replaced by $\sigma_{\mathrm{min}}$. This stems from the fact that the KL divergence between the distributions in equations (\ref{eqn:condn1_D}) and (\ref{eqn:condn2_D}) can be upper bounded by the smallest of value variance amongst all the noise entries.
\end{remark}

It is worth noting that our lower bounds on the minimax risk relate directly to the work in \cite{soni2014noisy}, which gives upper bounds for matrix completion problems under similar sparse factor models. The normalized (per-element) Frobenius error for the sparsity-penalized maximum likelihood estimator under a Gaussian noise model presented in \cite{soni2014noisy} satisfies
\begin{equation}\label{eqn:upper_bound}
  \cfrac{\EE_{\Yb_{\cS}}\left[\norm{\hat{\Xb}-\Xb^*}_{F}^2\right]}{n_1n_2}= \cO\left((\sigma\wedge\Amax)^2\bigg(\frac{n_1r+k}{m}\bigg)\log(n_1\vee n_2)\right).
\end{equation}
A comparison of (\ref{eqn:upper_bound}) to our results in \Cref{eqn:cor:AGN} imply that the rate attained by the estimator presented in \cite{soni2014noisy} is minimax optimal up to a logarithmic factor when there is (on an average), at least one non-zero element per columns of $\Ab^*$ (i.e. $k\geq n_2$), and provided we make sufficient observations (i.e $m\succeq n_1r+k$).

Another direct point of comparison to our result here is the low rank matrix completion problem with entry-wise observations considered in \cite{koltchinskii2011nuclear}. In particular, if we adopt the lower bounds obtained in Theorem 6 of that work to our settings, we observe that the risk involved in estimating rank-$r$ matrices that are sampled uniformly at random follows
\begin{eqnarray}\label{eqn:lowrank_result}
  \nonumber
  \cfrac{\EE_{\Yb_{\cS}}\left[\norm{\hat{\Xb}-\Xb^*}_{F}^2\right]}{n_1n_2} & = & \Omega\left((\sigma\wedge\Xmax)^2\bigg(\frac{(n_1\vee n_2)r}{m}\bigg)\right)\\
  & = & \Omega\left((\sigma\wedge\Xmax)^2\bigg(\frac{(n_1+n_2)r}{m}\bigg)\right),
\end{eqnarray}
where the last equality follows from the fact that $n_1\vee n_2 \geq (n_1+n_1)/2$. If we consider non-sparse factor models (where $k=rn_2$), it can be seen that the product $\Xb^*=\Db^*\Ab^*$ is low-rank with rank$(\Xb^*)\leq r$ and our problem reduces to the one considered in \cite{koltchinskii2011nuclear} (with $m\geq(n_1\vee n_2)r$, an assumption made in \cite{koltchinskii2011nuclear}). Under the conditions described above, and our assumption in Section \ref{subsec:obs_model} that $\Xmax=\Theta(\Amax)$, the lower bound given in (\ref{eqn:cor:AGN}) (or it's counterpart for the large sample regime) coincides with (\ref{eqn:lowrank_result}). However the introduction of sparsity brings additional structure which can be exploited in estimating the entries of $\Xb^*$, thus decreasing the risk involved.

\subsection{Additive Laplace Noise}\label{subsec:ALN}
  The following theorem gives a lower bound on the minimax risk in settings where the observations $\Yb_\cS$ are corrupted with heavier tailed noises; its proof is given in Appendix \ref{proof:cor:ALN}.

 \begin{corollary}[\bf{Lower bound for Laplacian Noise}]\label{cor:ALN}
   Suppose $Y_{i,j}=X^*_{i,j}+\xi_{i,j}$, where $\xi_{i,j}$ are i.i.d Laplace$(0,\tau)$, $\tau>0$, $\forall (i,j)\in\cS$. There exist absolute constants $C,\gamma>0$ such that for all $n_1, n_2\geq2$, $r\in[n_1 \wedge n_2]$, and $r\leq k\leq n_1n_2/2$, the minimax risk for sparse factor matrix completion over the model class $\cX(r,k,\Amax)$ obeys
   \begin{equation}\label{eqn:cor:ALN}
	\cR^*_{\cX(r,k,\Amax)}\geq C\cdot \min\Bigg\{\Delta(k,n_2)\mathrm{A^2_{max}},\gamma^2\:\tau^{-2}\bigg(\frac{n_1r+k}{m}\bigg)\!\Bigg\}.
  \end{equation}
 \end{corollary}

When we compare the lower bounds obtained under this noise model to the results of the previous case it can be readily seen that the overall error rates achieved are similar in both cases. Since we have the variance of Laplace($\tau$) random variable to be $(2/\tau^2)$, the leading term $\tau^{-2}$ in (\ref{eqn:cor:ALN}) is analogous to the $\sigma^2$ factor which appears in the error bound for Gaussian noise. Using (\ref{eqn:cor:ALN}), we can observe that the complexity penalized maximum likelihood estimator described in \cite{soni2014noisy} is minimax optimal up to a constant times a logarithmic factor, $\tau\Xmax\log(n_1\vee n_2)$ in the large sample regime, and when there is (on an average), at least one non-zero element per columns of $\Ab^*$ (i.e. $k\geq n_2$).

%======================================================================================================

\subsection{One-bit Observation Model}\label{subsec:1bit}
We consider here a scenario where the observations are quantized to a single bit i.e. the observations $Y_{i,j}$ can take only binary values (either $0$ or $1$). Quantized observation models arise in many collaborative filtering applications where the user ratings are quantized to fixed levels, in quantum physics, communication networks, etc. (see, e.g. discussions in \cite{davenport20141,haupt2014sparse}). %Another application of such models is to enforce global rate constraints in communication networks where transmitted data is quantized and sent to a centralized location for inference.

For a given sampling set $\cS$, we consider the observations $\Yb_{\cS}$ to be conditionally (on $\cS$) independent random quantities defined by
\begin{equation}\label{defn:1bitmodel}
Y_{i,j}= \mathbf{1}_{\{Z_{i,j}\geq0\}}, \quad (i,j)\in\cS,
\end{equation}
where
\begin{equation*}
	Z_{i,j}=\Xb^*_{i,j}-W_{i,j}.
\end{equation*}
Here the $\{W_{i,j}\}_{(i,j)\in\cS}$ are i.i.d continuous zero-mean scalar noises having (bounded) probability density function $f(w)$ and cumulative density function $F(w)$ for $w\in\RR$, and $\mathbf{1}_{\{\cA\}}$ is the indicator function which takes the value $1$ when the event $\cA$ occurs (or is true) and zero otherwise. Our observations are thus quantized, corrupted versions of the true underlying matrix entries. Note that the independence of $W_{i,j}$ implies that the elements $Y_{i,j}$ are also independent. Given this model, it can be easily seen that each $Y_{i,j},\ (i,j)\in\cS$, is a Bernoulli random variable whose parameter is a function of the true parameter $\Xb^*_{i,j}$, and the cumulative density function $F(\cdot)$. In particular, for any $(i,j)\in\cS$, we have $\text{Pr}(Y_{i,j}=1)=\text{Pr}(W_{i,j}\leq \Xb^*_{i,j})=F(\Xb^*_{i,j})$. Hence the joint pmf of the observations $\Yb_{\cS}\in\{0,1\}^{|\cS|}$ (conditioned on the underlying matrix entries) can be written as,
\begin{equation}\label{eqn:1bit_pmf}
p_{\Xb^*_{\cS}}(\Yb_{\cS})= \prod_{(i,j)\in\cS}[F(\Xb^*_{i,j})]^{Y_{i,j}}[1-F(\Xb^*_{i,j})]^{1-Y_{i,j}} \ .
\end{equation}
We will further assume that 
%$F(\cdot)$ is Lipschitz continuous with 
$F(r\Amax)<1 \text{ and } F(-r\Amax)>0$, which will allow us to avoid some pathological scenarios in our analyses. In such settings, the following theorem gives a lower bound on the minimax risk; its proof appears in Appendix \ref{proof:cor:1bit}.
\begin{corollary}[\bf{Lower bound for One-bit observation model}]\label{cor:1bit}
	Suppose that the observations $Y_{i,j}$ are obtained as described in (\ref{defn:1bitmodel}) where $W_{i,j}$ are i.i.d continuous zero-mean scalar random variables as described above, and define
	\begin{equation}\label{defn:cf_amax}
	c_{F,r\Amax}\triangleq\left(\sup\limits_{|t|\leq r\Amax}\cfrac{1}{F(t)(1-F(t))}\right)^{\!1/2}\!\left(\sup\limits_{|t|\leq r\Amax} f^2(t)\right)^{1/2} \ .
	\end{equation}
	There exist absolute constants $C,\gamma>0$ such that for all $n_1, n_2\geq2$, $1\leq r\leq (n_1 \wedge n_2)$, and $r\leq k\leq n_1n_2/2$, the minimax risk for sparse factor matrix completion over the model class $\cX(r,k,\Amax)$ obeys
	\begin{equation}\label{eqn:cor:1bit}
	\cR^*_{\cX(r,k,\Amax)}\geq C\cdot \min\Bigg\{\Delta(k,n_2)\mathrm{A^2_{max}},\gamma^2\:c_{F,r\Amax}^{-2}\bigg(\frac{n_1r+k}{m}\bigg)\!\Bigg\}.
	\end{equation}
\end{corollary}

It worth commenting on the relevance of our result (in the linear sparsity regime) to the upper bounds established in \cite{soni2014noisy}, for the matrix completion problem under similar settings. The normalized (per element) error of the complexity penalized maximum likelihood estimator described in \cite{soni2014noisy} obeys
\begin{equation}\label{eqn:upperbound:1bit}
\cfrac{\EE_{\Yb_{\cS}}\left[\norm{\hat{\Xb}-\Xb^*}_{F}^2\right]}{n_1n_2}=
\cO\left(\left(\frac{c^2_{F,r\Amax}}{c'_{F,r\Amax}}\right)\left(\frac{1}{c^2_{F,r\Amax}} +\mathrm{X^2_{max}}\right) \bigg(\frac{n_1r+k}{m}\bigg)\log(n_1\vee n_2)\right),
\end{equation}
where $\Xmax\ (\geq0)$ is the upper bound on the entries of the matrix to be estimated and $c'_{F,r\Amax}$ is defined as
\begin{equation}\label{def:c'f_amax}
c'_{F,r\Amax}\triangleq \inf\limits_{|t|\leq r\Amax}\cfrac{f^2(t)}{F(t)(1-F(t))}.
\end{equation}

Comparing (\ref{eqn:upperbound:1bit}) with the lower bound established in (\ref{eqn:cor:1bit}), we can see that estimator described in \cite{soni2014noisy} is minimax optimal up to a logarithmic factor (in the large sample regime, and with $k\geq n_2$) when the term $(c^2_{F,r\Amax}/c'_{F,r\Amax})$ is bounded above by a constant. The lower bounds obtained for the one-bit observation model and the Gaussian case essentially exhibit the same dependence on the matrix dimensions ($n_1,n_2\text{ and }r$), sparsity ($k$) and the nominal number of measurements ($m$), except for the leading term (which explicitly depends on the distribution of the noise variables $W_{i,j}$ for the one-bit case). Such a dependence in error rates between rate-constrained tasks and their Gaussian counterparts was observed in earlier works on rate-constrained parameter estimation \cite{ribeiro2006bandwidth,luo2005universal}.% and also in \cite{davenport20141} which analyzes low-rank matrix completion with single-bit measurements.

It is also interesting to compare our result with the lower bounds for the one-bit (low rank) matrix completion problem considered in \cite{davenport20141}. In that work, the authors establish that the risk involved in matrix completion over a (convex) set of max-norm and nuclear norm constrained matrices (with the decreasing noise pdf $f(t)$ for $t>0$) obeys
\begin{eqnarray}\label{eqn:1bit:lowrank}
\nonumber
\cfrac{\EE_{\Yb_{\cS}}\left[\norm{\hat{\Xb}-\Xb^*}_{F}^2\right]}{n_1n_2} &=& \Omega\left(\Xmax\sqrt{\frac{1}{c'_{F,r\Amax}}}\sqrt{\frac{(n_1\vee n_2)r}{m}}\right) \\
&=& \Omega\left(\Xmax\sqrt{\frac{1}{c'_{F,r\Amax}}}\sqrt{\frac{(n_1+n_2)r}{m}}\right),
\end{eqnarray}
where $c'_{F,r\Amax}$ is defined as in (\ref{def:c'f_amax}). As long as $c^2_{F,r\Amax}$ and $\sqrt{c'_{F,r\Amax}}$ are comparable, the leading terms of our bound and (\ref{eqn:1bit:lowrank}) are analogous to each other. In order to note the difference between this result and ours, we consider the case when $\Ab^*$ is not sparse i.e., we set $k=rn_2$ in (\ref{eqn:cor:1bit}) so that the resulting matrix $\Xb^*$ is low-rank (with rank$(\Xb)\leq r$). For such a setting, our error bound (\ref{eqn:cor:1bit}) scales in proportion to the ratio of the degrees of freedom $(n_1+n_2)r$ and the nominal number of observations $m$, while the bound in \cite{davenport20141} scales to the square root of that ratio. %Thus our bound appears to be tighter (up to leading constants) when $m\leq c\cdot r(n_1+n_2)$ for some constant $c>1$. In this context its worth noting that, while assumptions on monotonicity of the pdf $f(\cdot)$ were made in \cite{davenport20141}, we impose a much lesser constraint on the noise variables viz. Lipschitz continuity of $F(\cdot)$.

A more recent work \cite{lafond2014probabilistic}, proposed an estimator for the low-rank matrix completion on finite alphabets and establishes convergence rates faster than in \cite{davenport20141}. On casting their results to our settings, the estimation error in \cite{lafond2014probabilistic} was shown to obey
\begin{equation}\label{eqn:lafond:1bit}
\cfrac{\norm{\hat{\Xb}-\Xb^*}_{F}^2}{n_1n_2}=
\cO\left(\left(\frac{c^2_{F,r\Amax}}{c'_{F,r\Amax}}\right)^2 \bigg(\frac{(n_1+n_2)r}{m}\bigg)\log(n_1+n_2)\right).
\end{equation}
On comparing (\ref{eqn:lafond:1bit}) with the our lower bounds (for the low-rank case, where $k=rn_2$), it is worth noting that their estimator achieves minimax optimal rates up to a logarithmic factor when the ratio $(c^2_{F,r\Amax}/c'_{F,r\Amax})$ is bounded above by a constant.
%======================================================================================================

\subsection{Poisson-distributed Observations}\label{sec:poisson}
Let us now consider a scenario where the data maybe observed as discrete `counts' (which is common in imaging applications e.g., number of photons hitting the receiver per unit time). A popular model for such settings is the Poisson model, where all the entries of the matrix $\Xb^*$ to be estimated are positive and our observation $Y_{i,j}$ at each location $(i,j)\in\cS$ is an independent Poisson random variable with a rate parameter $\Xb^*_{i,j}$. The problem of matrix completion now involves the task of Poisson denoising. Unlike the previous cases, this problem cannot be directly cast into the setting of our general result, as there is an additional restriction on the model class that the entries of $\Xb^*$ are strictly bounded away from zero. A straightforward observation that follows is that the sparse factor $\Ab^*$ in the factorization cannot have any zero valued columns. Hence we have that $k\geq n_2$ be satisfied as a necessary (but not a sufficient) condition in this case. The approach we use to derive the following result is similar in spirit to the previous cases and is described in Appendix \ref{proof:thm:poisson}.

\begin{theorem}[\bf{Lower bound for Poisson noise}]\label{thm:poisson}
	Suppose that the entries of the matrix $\Xb^*$ satisfy $\min\limits_{i,j} X_{i,j}\geq \Xmin$ for some constant $0<\Xmin\leq\Amax$ and the observations $Y_{i,j}$ are independent Poisson distributed random variable with rates $X^*_{i,j}\ \forall(i,j)\in\cS$. There exist absolute constants $C,\gamma>0$ such that for all $n_1, n_2\geq2$, $r\in[n_1 \wedge n_2]$, and $n_2\leq k\leq n_1n_2/2$, the minimax risk for sparse factor matrix completion over the model class $\cX'(r,k,\Amax,\Xmin)$ which is a subset of $\cX(r,k,\Amax)$ comprised of matrices with positive entries, obeys
	\begin{equation}\label{eqn:thm:poisson}
	\cR^*_{\cX'(r,k,\Amax,\Xmin)} \geq C\cdot \min\Bigg\{ \tilde\Delta(k,n_2,\delta)\mathrm{A^2_{max}}  ,\gamma^2\:\Xmin\bigg(\frac{n_1r+k-n_2}{m}\bigg)\!\Bigg\}.
	\end{equation}
where $\delta = \frac{\Xmin}{\Amax}$, and the function $\tilde\Delta(k,n_2,\delta)$ is given by 
	\begin{equation}\label{def:Delta_poiss}
		\tilde\Delta(k,n_2,\delta) = \min\left\{(1-\delta)^2,\left(\frac{k-n_2}{n_2}\right)\right\}.
	\end{equation}
	
\end{theorem}

As in the previous cases, our analysis rests on establishing quadratic upper bounds on the KL divergence to obtain parametric error rates for the minimax risk; a similar approach was used in \cite{raginsky2010compressed}, which describes performance bounds on compressive sensing sparse signal estimation task under a Poisson noise model, and in \cite{kolaczyk2004multiscale}. Recall that the lower bounds for each of the preceding cases exhibited a leading factor to the parametric rate, which was essentially the noise variance. Note that for a Poisson observation model, the noise variance equals the rate parameter and hence depends on the true underlying matrix entry. So we might interpret the factor $\Xmin$ in (\ref{eqn:thm:poisson}) as the minimum variance of all the independent (but not necessarily identically distributed) Poisson observations and hence is somewhat analogous to the results presented for the Gaussian and Laplace noise models.

The dependence of the minimax risk on the nominal number of observations ($m$), matrix dimensions ($n_1,n_2,r$), and sparsity factor $k$, is encapsulated in the two terms, $\left(\frac{n_1r}{m}\right)$ and $\left(\frac{k-n_2}{m}\right)$. The first term, which corresponds to the error associated with the dictionary term $\Db^*$ is exactly the same as with the previous noise models. However we can see that the term associated with the sparse factor $\Ab^*$ is a bit different from the other models discussed. In a Poisson-distributed observation model, we have that the entries of the true underlying matrix to be estimated are positive (which also serves as the Poisson rate parameter to the observations $Y_{i,j}$). A necessary implication of this is that the sparse factor $\Ab^*$ should contain no zero-valued columns, or every columns should have at least one non-zero entry (and hence we have $k\geq n_2$). This reduces the effective number of \textit{degrees of freedom} (as described in Section \ref{subsec:AGN}) in the sparse factor from $k$ to $k-n_2$, thus reducing the overall minimax risk.

It is worth further commenting on the relevance of this result (in the large sample regime) to the work in \cite{soni2014noisy}, which establishes error bounds for Poisson denoising problems with sparse factor models. From Corollary III.3 of \cite{soni2014noisy}, we see that the normalized (per element) error of the complexity penalized maximum likelihood estimator obeys
\begin{equation}\label{eqn:upperbound:poisson}
\cfrac{\EE_{\Yb_{\cS}}\left[\norm{\hat{\Xb}-\Xb^*}_{F}^2\right]}{n_1n_2}=
\cO\left(\left(\Xmax +\frac{\Xmax}{\Xmin}\cdot\mathrm{X^2_{max}}\right) \bigg(\frac{n_1r+k}{m}\bigg)\log(n_1\vee n_2)\right),
\end{equation}
where $\Xmax$ is the upper bound on the entries of the matrix to be estimated. Comparing (\ref{eqn:upperbound:poisson}) with the lower bound established in (\ref{eqn:thm:poisson}) (or again, it's counterpart in the large sample regime), we can see that estimator described in \cite{soni2014noisy} is minimax optimal w.r.t to the matrix dimension parameters up to a logarithmic factor (neglecting the leading constants) when $k\geq 2n_2$. %However as far as the leading term is concerned, the upper bounds established in \cite{soni2014noisy} diverge (tend to $\infty$) as $\Xmin$ tends to $0$, while the lower bounds we establish behave differently.

We comment a bit on our assumption that the elements of the true underlying matrix $\Xb^*$, be greater than or equal to some $\Xmin>0$. %; similar minimum rate assumptions were made in \cite{raginsky2010compressed,kolaczyk2004multiscale}.
Here, this parameter shows up as a multiplicative term on the parametric rate $(\frac{n_1r+k-n_2}{m})$, which suggests that the minimax risk vanishes to 0 at the rate of $\Xmin/m$ (when the problem dimensions are fixed). This implication is in agreement with the classical Cram\'{e}r-Rao lower bounds which states that the variance associated with estimating a Poisson($\theta$) random variable using $m$ iid observations decays at the rate $\theta/m$ (and achieved by a sample average estimator). Thus our notion that the denoising problem becomes \textit{easier} as the rate parameter decreases is intuitive and is consistent with classical analyses. On this note, we briefly mention recent efforts which do not make assumptions on the minimum rate of the underlying Poisson processes; for matrix estimation tasks as here \cite{soni2014estimation}, and for sparse vector estimation from Poisson-distributed compressive observations \cite{jiang2014minimax}. %The ideas presented in these works can be adopted to the problem considered here, but involve imposing different (maybe even stronger) incoherence assumptions on $\Xb^*$ than the bounded entry model defined here.

% ======================================================================

\section{Conclusion}\label{sec:concl}

In this paper, we established minimax lower bounds for sparse factor matrix completion tasks, under very general noise/corruption models.  We also provide lower bounds for several specific noise distributions that fall under our general noise model. This indicates that property (\ref{eqn:quad_KLdiv_bound}), which requires that the scalar KL divergences of the noise distribution admit a quadratic upper bounded in terms of the underlying parameters, is not overly restrictive in many interesting scenarios. 

%Although closed form expressions for lower bounds under generic models have not been discussed here, we would like to point out that for noise models whose scalar KL divergence admits an upper bound quadratic in the difference of the underlying matrix elements, the minimax risk obeys the bounds established in Section \ref{sec:result}.

A unique aspect of our analysis is its applicability to matrices representable as a product of structured factors.  While our focus here was specifically on models in which one factor is sparse, the approach we utilize here to construct packing sets extends naturally to other structured factor models (of which standard low-rank models are one particular case).  A similar analysis to that utilized here could also be used to establish lower bounds on estimation of structured tensors, for example, those expressible in a Tucker decomposition with sparse core, and possibly structured factor matrices (see, e.g., \cite{kolda2009tensor} for a discussion of Tucker models). We defer investigations along these lines to a future effort. 

\section{Acknowledgements}\label{sec:ack}
The authors are grateful to the anonymous reviewer for their detailed and thorough evaluations of the paper. In particular, we thank the reviewer for pointing out some subtle errors in the initial versions of our main results that motivated us to obtain tighter lower bounds. 

%Our work here provides useful theoretical tools to establish minimax bounds on a wide range of compressive sensing and matrix completion problems. In future work, we could possibly consider more generalized non-uniform sampling schemes.

%The \Cref{thm:AGN,thm:ALN,thm:poisson,thm:1bit} give valid lower bounds on a very broad framework of sparse factor models. To the best of our knowledge, lower bounds tighter than that presented in the above results have not been established.  

\appendix

In order to prove Theorem \ref{thm:general} we use standard minimax analysis techniques, namely the following theorem (whose proof is available in \cite{tsybakov2008introduction}),
\begin{theorem}[Adopted from Theorem 2.5 in \cite{tsybakov2008introduction}]
\label{thm:tsybakov}
  Assume that $M\geq2$ and suppose that there exists a set with finite elements, $\cX=\ \{\Xb_0, \Xb_{1}, \dots \Xb_{M}\}\subset \cX(r,k,\Amax)$ such that
  \begin{itemize}
    \item $d(\Xb_{j},\Xb_{k})\geq2s,\quad \forall\, 0\leq j<k\leq M;\,$ where $d(\cdot,\cdot):\cX\times\cX\to \RR$ is a semi-distance function, and
    \item $\frac{1}{M}\sum\limits_{j=1}^{M}K(\PP_{\Xb_j},\PP_{\Xb_0})\leq \alpha\log M$ with $0<\alpha<1/8$.
  \end{itemize}
  Then
  \begin{eqnarray}\label{eqn:tsybakov}
    \nonumber % Remove numbering (before each equation)
    \inf\limits_{\hat{\Xb}}\sup\limits_{\Xb\in\cX(r,k,\Amax)} \PP_{\Xb}(d(\hat{\Xb},\Xb)\geq s) &\geq& \inf\limits_{\hat{\Xb}}\sup\limits_{\Xb\in\cX} \PP_{\Xb}(d(\hat{\Xb},\Xb)\geq s) \\
    &\geq& \frac{\sqrt{M}}{1+\sqrt{M}} \left(1-2\alpha-\sqrt{\frac{2\alpha}{\log M}}\right)>0.
  \end{eqnarray}
\end{theorem}
Here the first inequality arises from the fact that the supremum over a class of matrices $\cX$ is upper bounded by that of a larger class $\cX(r,k,\Amax)$ (or in other words, estimating the matrix over an uncountably infinite class is at least as difficult as solving the problem over any finite subclass). We thus reduce the problem of matrix completion over an uncountably infinite set $\cX(r,k,\Amax)$, to a carefully chosen finite collection of matrices $\cX \subset \cX(r,k,\Amax)$ and lower bound the latter which then gives a valid bound for the overall problem.

In order to obtain tight lower bounds, it essential to carefully construct the class $\cX$ (which is also commonly called a \emph{packing set}) with a large cardinality, such that its elements are also as far apart as possible in terms of (normalized) Frobenius distance, which is our choice of semi-distance metric.

\subsection{Proof of Theorem \ref{thm:general}}\label{proof:thm:general}
Let us define a class of matrices $\cX\subset\RR^{n_1\times n_2}$ as
\begin{equation}\label{defn:reduction_class}
  \cX \triangleq \{\Xb=\Db\Ab:\:\Db\in\cD,\:\Ab\in\cA\},
\end{equation}
where the factor classes $\cD\subset\RR^{n_1\times r}$ and $\cA\subset\RR^{r\times n_2}$ are constructed as follows for $\gamma_d,\, \gamma_a\leq1$ (to be quantified later)
\begin{equation}\label{defn:calD}
  \cD \triangleq \left\{\Db\in \RR^{n_1\times r}: \:\Db_{i,j}\in\left\{0,1,d_0\right\}, \forall (i,j)\in[n_1]\times[r]\right\},
\end{equation}
and
\begin{equation}\label{defn:calA}
  \cA \triangleq \left\{\Ab\in \RR^{r\times n_2}: \:\Ab_{i,j}\in\{0,\Amax,a_0\}, \forall (i,j)\in[r]\times[n_2],\  \norm{\Ab}_0\leq k\right\},
\end{equation}
where $d_0 \triangleq \min\left\{1,\frac{\gamma_d\cdot\mu_D}{\Amax\sqrt{\Delta(k,n_2)}} \left(\frac{n_1r}{m}\right)^{1/2}\right\},$ $a_0 \triangleq \min\left\{\Amax,\frac{\gamma_a\cdot\mu_D}{\sqrt{\Delta(k,n_2)}}\left(\frac{k}{m}\right)^{1/2}\right\}$, and $\Delta(k,n_2)$ is defined in (\ref{eqn:Delta}). 

Clearly $\cX$ as defined in (\ref{defn:reduction_class}) is a finite class of matrices which admits a factorization as in Section \ref{subsec:obs_model}, so $\cX\subset\cX(r,k,\Amax)$. We consider the lower bounds involving the non-sparse term, $\Db$ and the sparse factor $\Ab$ separately and then combine those results to a get an overall lower bound on the minimax risk $\cR^*_{\cX(r,k,\Amax)}$.

Let us first establish the lower bound obtained by using the sparse factor $\Ab$. In order to do this, we define a set of sparse matrices $\bar{\cA}\subset\cA$, where all the nonzero terms are stacked in the first $r'=\left\lceil\frac{k}{n_2}\right\rceil$ rows. Formally we define 
\begin{equation}\label{def:calAbar}
\bar\cA\triangleq\left\{\Ab\in\RR^{r\times n_2}:\,\Ab =\left(\Ab_{nz}|\mathbf{0}_A\right)^T, (\Ab_{nz})_{i,j}\in\{0,a_0\},\forall(i,j)\in[n_2]\times [r'], \norm{\Ab_{nz}}_0\leq k\right\},
\end{equation}
where $\Ab_{nz}$ is an $n_2\times r'$ sparse matrix with at most $k$ non zeros and $\mathbf{0}_A$ is an $n_2\times(r-r')$ zero matrix. Let us now define the finite class of matrices $\cX_{A}\subset\cX$ as
\begin{equation}\label{def:calX_A}
\cX_{A}\triangleq \left\{\Xb=\Db_I\Ab:\: \Ab\in\bar\cA \right\},
\end{equation}
where $\Db_I$ is made up of block zeros and block identity matrices, and defined as follows 
\begin{equation}\label{def:D_I}
\Db_I \triangleq \left(
\begin{array}{c|c}
\Ib_{r'} & \mathbf{0} \\
\hline
\vdots & \vdots \\
\hline
\Ib_{r'} & \mathbf{0} \\
\hline
\mathbf{0} & \mathbf{0}
\end{array}
\right),
\end{equation}
where, $\Ib_{r'}$ denotes the $r'\times r'$ identity matrix.

The definitions in \cref{def:calAbar,def:calX_A,def:D_I} imply that the elements of $\cX_A$ form block matrices which are of the form $\left(\begin{array}{c|c|c|c}
            \Ab_{nz}&\cdots&\Ab_{nz}&\mathbf{0}
            \end{array}\right)^T$, with 
$\lfloor n_1/r'\rfloor$ blocks of $\Ab_{nz}$, $\forall \Ab\in\bar\cA$ and the rest is a zero matrix of the appropriate dimension. Since the entries of $\Ab_{nz}$ can take only one of two values 0 or $a_0$, and since there are at most $k$ non-zero elements (due to the sparsity constraint), the Varshamov-Gilbert bound (cf. Lemma 2.9 in \cite{tsybakov2008introduction}) guarantees the existence of a subset $\cX^0_A\subseteq\cX_A$ with cardinality Card($\cX^0_A$) $\geq2^{k/8}+1$, containing the $n_1\times n_2$ zero matrix $\mathbf{0}$, such that for any 2 distinct elements $\Xb_1,\Xb_2\in\cX^0_A$ we have,
\begin{eqnarray}\label{eqn:condn1_A}
  \nonumber 
  \norm{\Xb_1-\Xb_2}_F^2 
  &\geq&\left(\frac{k}{8}\right)\left\lfloor\frac{n_1}{r'}\right\rfloor a^2_0\\ 
  \nonumber
  &=&\left(\frac{k}{8}\right)\left\lfloor\frac{n_1}{\lceil k/n_2\rceil}\right\rfloor \min\left\{\mathrm{A^2_{max}},\frac{\gamma_a^2\mu_D^2}{\Delta(k,n_2)}\left(\frac{k}{m}\right)\right\}\\
  \nonumber
  &\geq&\left(\frac{n_1n_2}{32}\right)\underbrace{\left(\frac{k\wedge n_2}{n_2}\right)}_{=\Delta(k,n_2)} \min\left\{\mathrm{A^2_{max}},\frac{\gamma_a^2\mu_D^2}{\Delta(k,n_2)}\left(\frac{k}{m}\right)\right\}\\
  &=& \left(\frac{n_1n_2}{32}\right) \min\left\{\Delta(k,n_2)\mathrm{A^2_{max}},\gamma_a^2\mu_D^2\left(\frac{k}{m}\right)\right\},
\end{eqnarray}
where the second to last inequality comes from the fact that $k\leq (n_1n_2)/2$, and $\lfloor x\rfloor \geq x/2 \; \forall x\geq1$.

For any $\Xb\in\cX^0_A$, consider the KL divergence of $\PP_\mathbf{0}$ from $\PP_\Xb$,
\begin{eqnarray}
 \nonumber
	K(\PP_\Xb,\PP_\mathbf{0})&=& \EE_\Xb \left[\log \frac{p_{\Xb_{\cS}}(\Yb_\cS)}{p_{\mathbf{0}}(\Yb_\cS)}\right]\\
\label{eqn:KLdiv_A:1} 
	&=& \sum_{i,j} K(\PP_{X_{i,j}},\PP_{\mathbf{0}_{i,j}}) \cdot \frac{m}{n_1n_2} \\
\label{eqn:KLdiv_A:2}
	&\leq& \frac{m}{n_1n_2}\left(\frac{1}{2\mu_D^2}\right)\sum_{i,j} |X_{i,j}|^2, 
\end{eqnarray}
where (\ref{eqn:KLdiv_A:1}) is obtained by conditioning\footnote{Here, both the observations $\Yb_{\cS}$, and the sampling set $\cS$ are random quantities. Thus by conditioning w.r.t to $\cS$, we get $\EE_{\Xb}\triangleq\EE_{\Yb_{\cS}\sim p_{\Xb_\cS}}\left[\log \Big(p_{\Xb_{\cS}}(\Yb_\cS)/p_{\mathbf{0}}(\Yb_\cS)\Big)\right] = \EE_{\cS}\left[\EE_{\Xb_{\cS}|\cS}\left[\log \Big(p_{\Xb_{\cS}}(\Yb_\cS)/p_{\mathbf{0}}(\Yb_\cS)\Big)\right] \right]$. Since $\cS$ is generated according to the independent Bernoulli$(m/n_1n_2)$ model, $\EE_{\cS}[\cdot]$ yields the constant term $\frac{m}{n_1n_2}$. We shall use such conditioning techniques in subsequent proofs as well.}
%\footnotemark -------------------------------
the expectation w.r.t the sampling set $\cS$, and (\ref{eqn:KLdiv_A:2}) follows from the assumption on noise model (\ref{eqn:quad_KLdiv_bound}). To further upper bound the RHS of (\ref{eqn:KLdiv_A:2}), we note that the maximum number of nonzero entries in any $\Xb\in\cX^0_A$ is at most $n_1(k\wedge n_2)$ by construction of the sets $\cX_A$ and $\bar{\cA}$ in (\ref{def:calX_A}), (\ref{def:calAbar}) respectively. Hence we have
\begin{eqnarray}
\nonumber
K(\PP_\Xb,\PP_\mathbf{0}) &\leq& \frac{m}{2\mu_D^2}\underbrace{\left(\frac{k\wedge n_2}{n_2}\right)}_{=\Delta(k,n_2)}a_0^2\\
\label{eqn:KLdiv_A:3} 
&=& \frac{m}{2\mu_D^2} \min\left\{\Delta(k,n_2)\mathrm{A^2_{max}},\gamma_a^2\cdot\mu^2_D\left(\frac{k}{m}\right)\right\}.
\end{eqnarray}
From (\ref{eqn:KLdiv_A:3}) we see that
\begin{equation}\label{eqn:condn2_A}
  \frac{1}{\text{Card}(\cX^{0}_{A}) -1}\sum_{\Xb\in\cX^0_A} K(\PP_\Xb,\PP_\mathbf{0}) \leq\alpha\log(\text{Card}(\cX^0_A)-1),
\end{equation}
is satisfied for any $0<\alpha<1/8$ by choosing $0<\gamma_a<\frac{\sqrt{\alpha\log2}}{2}$. Equations (\ref{eqn:condn1_A}) and (\ref{eqn:condn2_A}) imply we can apply Theorem \ref{thm:tsybakov} (where the Frobenius error has been as used the semi-distance function) to yield
\begin{equation}\label{eqn:result_A}
  \inf\limits_{\hat{\Xb}}\sup\limits_{\Xb^*\in\cX_A} \PP_{\Xb^*}\left(\frac{\norm{\hat\Xb-\Xb^*}_F^2}{n_1n_2}\geq \frac{1}{64}\cdot \min\left\{\Delta(k,n_2)\mathrm{A^2_{max}},\gamma_a^2\mu_D^2\left(\frac{k}{m}\right)\!\right\} \right)\geq \beta,
\end{equation}
for some absolute constant $\beta\in(0,1)$.
%---------- Conditional expectation Foot Note---------------
%\footnotetext{
%Here, both the observations $\Yb_{\cS}$, and the sampling set $\cS$ are random quantities. Thus by conditioning w.r.t to $\cS$, we get $\EE_{\Xb}\triangleq\EE_{\Yb_{\cS}\sim p_{\Xb_\cS}}[\cdot]=\EE_{\cS}\left[\EE_{\Xb_{\cS}|\cS}[\cdot] \right]$. Since $\cS$ is generated according to the independent Bernoulli$(m/n_1n_2)$ model, $\EE_{\cS}[\cdot]$ yields the constant term $\frac{m}{n_1n_2}$. We shall use such conditioning techniques in subsequent proofs as well.
%}
%---------------------------------------------------------------------------------------------------------------------

We now consider the non-sparse factor $\Db$ to construct a testing set and establish lower bounds similar to the previous case. Let us define a finite class of matrices $\cX_D\subseteq\cX$ as
\begin{equation}\label{defn:calX_D}
  \cX_{D}\triangleq \left\{\Xb=\Db\Ab:\: \Db\in\bar{\cD}, \, \Ab=\Amax\left(\begin{array}{c|c|c|c}
                                          \Ib_r&\cdots&\Ib_r&\mathbf{0}_D
                                          \end{array}\right) \in\cA, \right\},
\end{equation}
where $\Ab$ is constructed with $\lfloor(k\wedge n_2)/r\rfloor$ blocks of $r\times r$ identity matrices (denoted by $\Ib_r$), $\mathbf{0}_D$ is $r\times \left(n_2 - \left\lfloor\frac{k\wedge n_2}{r}\right\rfloor r\right)$ zero matrix, and $\bar{\cD}\subseteq\cD$ is defined as
\begin{equation}\label{eqn:defn_calDbar}
  \bar{\cD}\triangleq\left\{\Db\in\RR^{n_1\times r}:\, \Db_{ij}\in\left\{0,d_0\right\},\, \forall (i,j)\in[n_1]\times[r]\right\}.
\end{equation}

The definition in (\ref{defn:calX_D}) is similar to that we used to construct $\cX_A$ and hence it results in a block matrix structure for the elements in $\cX_D$. We note here that there are $n_1r$ elements in each block $\Db$, where each entry can be either 0 or $d_0$. Hence the Varshamov-Gilbert bound (cf. Lemma 2.9 in \cite{tsybakov2008introduction}) guarantees the existence of a subset $\cX^0_D\subseteq\cX_D$ with cardinality Card($\cX^0_D$) $\geq2^{n_1r/8}+1$, containing the $n_1\times n_2$ zero matrix $\mathbf{0}$, such that for any 2 distinct elements $\Xb_1,\Xb_2\in\cX^0_D$ we have
\begin{eqnarray}
  \nonumber
  \norm{\Xb_1-\Xb_2}_F^2 
  &\geq& \left(\frac{n_1r}{8}\right)\left\lfloor\frac{k\wedge n_2}{r}\right\rfloor \mathrm{A^2_{max}}d_0^2 \\
\label{eqn:condn1_D}  
  &\geq& \left(\frac{n_1n_2}{16}\right) \min\left\{\mathrm{A^2_{max}}\Delta(k,n_2),\gamma^2_d\mu^2_D\left(\frac{n_1r}{m}\right)\right\},
\end{eqnarray}
where we use the fact that $(k\wedge n_2) = n_2\cdot\Delta(k,n_2)$, and $\lfloor x\rfloor \geq x/2 \; \forall x\geq1$ to obtain the last inequality.

%Using the expression for the joint pdf in (\ref{eqn:AGN_pdf}), for any $\Xb\in\cX^0_D$ and a sampling set $\cS$, the KL divergence of $\PP_\mathbf{0}$ from $\PP_\Xb$ satisfies

For any $\Xb\in\cX^0_D$, consider the KL divergence of $\PP_\mathbf{0}$ from $\PP_\Xb$,
\begin{eqnarray}
\nonumber
K(\PP_\Xb,\PP_\mathbf{0})
&=& \EE_\Xb \left[\log \frac{p_{\Xb_{\cS}}(\Yb_\cS)}{p_{\mathbf{0}}(\Yb_\cS)}\right]\\
\label{eqn:KLdiv_D:2}
&\leq& \frac{m}{n_1n_2}\left(\frac{1}{2\mu_D^2}\right)\sum_{i,j} |X_{i,j}|^2,
\end{eqnarray}
where the inequality follows from the assumption on noise model (\ref{eqn:quad_KLdiv_bound}). To further upper bound the RHS of (\ref{eqn:KLdiv_D:2}), we note that the maximum number of nonzero entries in any $\Xb\in\cX^0_D$ is at most $n_1(k\wedge n_2)$ by construction of the class $\cX_D$ in (\ref{defn:calX_D}). Hence we have
\begin{eqnarray}
K(\PP_\Xb,\PP_\mathbf{0})
\nonumber
&\leq& \frac{m}{2\mu_D^2}\underbrace{\left(\frac{k\wedge n_2}{n_2}\right)}_{=\Delta(k,n_2)}\mathrm{A^2_{max}}d_0^2\\
\label{eqn:KLdiv_D:3} 
&=& \frac{m}{2\mu_D^2} \min\left\{\Delta(k,n_2)\mathrm{A^2_{max}},\gamma_d^2\mu^2_D\left(\frac{n_1r}{m}\right)\right\}
\end{eqnarray}

From (\ref{eqn:KLdiv_D:3}) we see that %the following condition is satisfied
\begin{equation}\label{eqn:condn2_D}
  \frac{1}{\text{Card}(\cX^{0}_{D}) -1}\sum_{\Xb\in\cX^0_D} K(\PP_\Xb,\PP_\mathbf{0}) \leq\alpha'\log(\text{Card}(\cX^0_D)-1),
\end{equation}
is satisfied for any $0<\alpha'<1/8$ by choosing $0<\gamma_d<\frac{\sqrt{\alpha'\log2}}{2}$. Equations (\ref{eqn:condn1_D}) and (\ref{eqn:condn2_D}) imply we can apply Theorem \ref{thm:tsybakov} (where the Frobenius error has been as used the semi-distance function) to yield
\begin{equation}\label{eqn:result_D}
  \inf\limits_{\hat{\Xb}}\sup\limits_{\Xb^*\in\cX_D} \PP_{\Xb^*}\left(\frac{\norm{\hat\Xb-\Xb^*}_F^2}{n_1n_2}\geq \frac{1}{64}\cdot \min\left\{\Delta(k,n_2)\mathrm{A^2_{max}},\gamma_d^2\mu_D^2\bigg(\frac{n_1r}{m}\bigg)\!\right\} \right)\geq \beta',
\end{equation}
for some absolute constant $\beta'\in(0,1)$. Inequalities (\ref{eqn:result_A}) and (\ref{eqn:result_D}) imply the result,
\begin{equation}\label{eqn:result_X}
  \inf\limits_{\hat{\Xb}}\sup\limits_{\Xb^*\in\cX(r,k,\Amax)} \PP_{\Xb^*}\left(\frac{\norm{\hat\Xb-\Xb^*}_F^2}{n_1n_2}\geq \frac{1}{128}\cdot\min\left\{\Delta(k,n_2)\mathrm{A^2_{max}},\gamma_D^2\:\mu_D^2\left(\frac{n_1r+k}{m}\right)\!\right\} \right)\geq (\beta'\wedge \beta),
\end{equation}
where $\gamma_D=(\gamma_d\wedge\gamma_a)$, is a suitable value for the leading constant, and we have $(\beta'\wedge \beta)\in(0,1)$. In order to obtain this result for the entire class $\cX(r,k,\Amax)$, we use the fact that solving the matrix completion problem described in Section \ref{subsec:obs_model} over a larger (and possibly uncountable) class of matrices is at least as difficult as solving the same problem over a smaller (and possibly finite) subclass. Applying Markov's inequality to (\ref{eqn:result_X}) directly yields the result of Theorem \ref{thm:general},  completing the proof. \hfill \qedsymbol
%==========================================================================================================
\subsection{Proof of Corollary \ref{cor:AGN}}\label{proof:cor:AGN}
For a Gaussian distribution with mean $x\in\RR$ and variance $\sigma^2$, denoted by $\PP_x \sim \cN(x,\sigma^2)$, we have 
\begin{equation}\label{eqn:AGN_pdf}
  p_x(z)=\frac{1}{\sqrt{2\pi\sigma^2}} \exp\left(-\frac{1}{2\sigma^2} (z-x)^2\right),\quad \forall z\in\RR.
\end{equation}
Using the expression for the pdf of a Gaussian random variable in (\ref{eqn:AGN_pdf}), the KL divergence of $\PP_x$ from $\PP_y$ (for any $y\in\RR$) satisfies
\begin{eqnarray}\label{eqn:KLdiv:scalar_AGN}
\nonumber % Remove numbering (before each equation)
	K(\PP_x,\PP_y) &=& \EE_x \left[\log \frac{p_x(z)}{p_y(z)}\right] \\
\label{eqn:AGN:KL_bound}
	&=&  \frac{1}{2\sigma^2}(x-y)^2.
\end{eqnarray}
The expression for KL divergence between scalar Gaussian distributions (with identical variances) given in (\ref{eqn:AGN:KL_bound}), obeys the condition (\ref{eqn:quad_KLdiv_bound}) with equality, where $\mu_D=\sigma$. Hence we directly appeal to Theorem \ref{thm:general} to yield the desired result.

\subsection{Proof of Corollary \ref{cor:ALN}}\label{proof:cor:ALN}

For a Laplace distribution with parameter $\tau>0$ centered at $x$, denoted $\PP_x\sim\text{Laplace}(x,\tau)$ where $x\in\RR$, the KL divergence of $\PP_x$ from $\PP_y$ (for any $y\in\RR$) can be computed by (relatively) straightforward calculation as
\begin{eqnarray}\label{eqn:KLdiv:scalar_ALN}
   \nonumber % Remove numbering (before each equation)
    K(\PP_x,\PP_y) &=& \EE_x \left[\log \frac{p_x(z)}{p_y(z)}\right] \\
     &=& \tau|x-y|-\big(1-e^{-\tau|x-y|}\big).
\end{eqnarray}
  Using a series expansion of the exponent in (\ref{eqn:KLdiv:scalar_ALN}) we have
\begin{eqnarray}\label{eqn:scalar_ALN:ineq}
  \nonumber % Remove numbering (before each equation)
     e^{-\tau|x-y|}&=&1-\tau|x-y|+\frac{(\tau|x-y|)^2}{2!}-\frac{(\tau|x-y|)^3}{3!}+\cdots \\
     &\leq& 1-\tau|x-y|+\frac{(\tau|x-y|)^2}{2!}.
\end{eqnarray}
Rearranging the terms in (\ref{eqn:scalar_ALN:ineq}) yields the result,
\begin{equation}\label{eqn:ALN:KL_bound}
K(\PP_x,\PP_y)\leq \frac{\tau^2}{2}(x-y)^2.
\end{equation}

With the upper bound on the KL divergence established in (\ref{eqn:ALN:KL_bound}), we directly appeal to Theorem \ref{thm:general} with $\mu_D = \tau^{-1}$  to yield the desired result. \hfill \qedsymbol
%======================================================================================================

\subsection{Proof of Corollary \ref{cor:1bit}}\label{proof:cor:1bit}
For any $\Xb,\ \Xb^*\in\cX(r,k,\Amax)$ using the pdf model described in (\ref{eqn:1bit_pmf}), it is straightforward to show that the scalar KL divergence is given by
\begin{equation}\label{eqn:KLdiv:1bit}
K(\PP_{X^*_{i,j}},\PP_{X_{i,j}})=
F(X^*_{i,j})\log\left(\frac{F(X^*_{i,j})}{F(X_{i,j})}\right)+(1-F(X^*_{i,j}))\log\left(\frac{1-F(X^*_{i,j})}{1-F(X_{i,j})}\right)
\end{equation}
for any $(i,j)\in\cS$. We directly use an intermediate result from \cite{soni2014noisy} to invoke a quadratic upper bound for the KL divergence term,
\begin{equation}\label{eqn:KLdiv:1bit:upperbound}
K(\PP_{X^*_{i,j}},\PP_{X_{i,j}})\leq \frac{1}{2}c^2_{F,r\Amax}(X^*_{i,j}-X_{i,j})^2,
\end{equation}
where $c_{F,r\Amax}$ is defined in (\ref{defn:cf_amax}). Such an upper bound in terms of the underlying matrix entries can be attained by following a procedure illustrated in \cite{haupt2014sparse}, where one first establishes quadratic bounds on the KL divergence in terms of the Bernoulli parameters, and then subsequently establishes a bound on the squared difference between Bernoulli parameters in terms of the squared difference of the underlying matrix elements.

With a quadratic upper bound on the scalar KL divergences in terms of the underlying matrix entries (\ref{eqn:KLdiv:1bit:upperbound}), we directly appeal to the result of Theorem \ref{thm:general} with $\mu_D = c_{F,r\Amax}^{-1} $ to yield the desired result. \hfill \qedsymbol

%======================================================================================================
\subsection{Proof of Theorem \ref{thm:poisson}}\label{proof:thm:poisson}
The Poisson observation model considered here assumes that all the entries of the underlying matrix $\Xb^*$ are strictly non-zero. We will use similar techniques as in Appendix \ref{proof:thm:general} to derive the result for this model. However we need to be careful while constructing the sample class of matrices as we need to ensure that all the entries of the members should be strictly bounded away from zero (and in fact $\geq \Xmin$). In this proof sketch, we will show how an appropriate packing set can be constructed for this problem, and obtain lower bounds using arguments as in Appendix \ref{proof:thm:general}.

As before, let us fix $\Db$ and establish the lower bounds due to the sparse factor $\Ab$ alone. For $\gamma_d,\, \gamma_a\leq1$ (which we shall qualify later), we construct the factor classes $\cD\subset\RR^{n_1\times r}$, and $\cA\subset\RR^{r\times n_2}$ 
\begin{equation}\label{defn:poiss:calD}
\cD \triangleq \left\{\Db\in \RR^{n_1\times r}: \:\Db_{i,j}\in\left\{0,1,\delta,d_0\right\}, \forall (i,j)\in[n_1]\times[r]\right\},
\end{equation}
and
\begin{equation}\label{defn:poiss:calA}
\cA \triangleq \left\{\Ab\in \RR^{r\times n_2}: \:\Ab_{i,j}\in\{0,\Xmin,\Amax,a_0\}, \forall (i,j)\in[r]\times[n_2],\ \norm{\Ab}_0\leq k\right\},
\end{equation}
where $\delta \triangleq \frac{\Xmin}{\Amax},\ d_0 \triangleq \min\left\{1-\delta, \frac{\gamma_d\cdot\sqrt{\Xmin}}{\Amax} \left(\frac{n_1r}{m}\right)^{1/2}\right\}$, $a_0 \triangleq \min\left\{\Amax, \gamma_a\cdot\sqrt{\frac{\Xmin}{\Delta(k-n_2,n_2)}}\left(\frac{k-n_2}{m}\right)^{1/2}\right\}$, and $\Delta(\cdot,\cdot)$ is defined in (\ref{eqn:Delta}). Similar to the previous case, we consider a subclass $\bar{\cA}\subset\cA$ with at most $k$ nonzero entries which are all stacked in the first $r'+1=\left\lceil\frac{k}{n_2}\right\rceil$ rows such that $\forall \Ab\in\bar\cA$ we have,
\begin{equation}\label{eqn:A:poisson}
(\Ab)_{ij}=\left\{
\begin{array}{cl}
\Xmin & \mbox{for }  i=1, j\in[n_2] \\
(\Ab)_{ij}\in \{0,a_0\} & \mbox{for } 2\leq i \leq r'+1, j\in[n_2] \\
0 & \mbox{otherwise}
\end{array}
\right. \ .
\end{equation}

Now we define a finite class of matrices $\cX_A\subset\RR^{n_1\times n_2}$ as
\begin{equation}\label{def:poiss:calX_A}
\cX_{A}\triangleq \left\{\Xb=(\Db_0+\Db_I)\Ab:\: \Ab\in\bar\cA \right\},
\end{equation}
where $\Db_I,\Db_0\in\cD$ are defined as
\begin{eqnarray}
\label{def:poiss:D_I}
\Db_I &\triangleq& \left(
\begin{array}{c|c|c}
\mathbf{0}_{r'} & \Ib_{r'} & \mathbf{0} \\
\hline
\vdots & \vdots \\
\hline
\mathbf{0}_{r'} & \Ib_{r'} & \mathbf{0} \\
\hline
\mathbf{0} & \mathbf{0} & \mathbf{0}
\end{array}
\right)\quad \& \\
\label{def:poiss:D_0}
\Db_0 &\triangleq& \big(\mathbf{1}_{n_1} \vert\: \mathbf{0}\big),
\end{eqnarray}
where $\mathbf{1}_{n_1}$ is the $n_1\times 1$ vector of all ones, and we have $\left\lfloor\frac{n_1}{r'}\right\rfloor$ blocks of $\mathbf{0}_{r'}$ (which is the $r'\times 1$ zero vector) and $\Ib_{r'}$ (which is the $r'\times r'$ identity matrix), in $\Db_I$.

The above definitions ensure that $\cX_A\subset\cX'(r,k,\Amax,\Xmin)$. In particular, for any $\Xb\in\cX_A$ we have
\begin{eqnarray}
	\nonumber
	\Xb 
	&=& (\Db_0+\Db_I)\Ab \\
	\label{eqn:poiss:classform}
	&=& \underbrace{\Xmin\mathbf{1}_{n_1}\cdot\mathbf{1}_{n_2}^T}_{=\Xb_0} + \Db_I\Ab',
\end{eqnarray}
where $(\Db_0+\Db_I)\in\cD$, $\Ab\in\bar\cA$, and $\Ab'\in\RR^{r\times n_2}$ just retains the rows 2 to $r'+1$, of $\Ab$. It is also worth noting here that the matrix $\Db_I\Ab'$ has $\left\lfloor\frac{n_1}{r'}\right\rfloor$ copies of the nonzero elements of $\Ab'$. Now let us consider the Frobenius distance %(Frobenius norm of the difference) 
between any two distinct elements $\Xb_1,\Xb_2\in\cX_A$,
  \begin{eqnarray}\label{eqn:poiss:frobeniusDist}
    \nonumber
    \norm{\Xb_1-\Xb_2}_F^2 &=& \norm{(\Db_0+\Db_{I})\Ab_1-(\Db_0+\Db_{I})\Ab_2}_F^2 \\
    \nonumber &=& \norm{\Xb_0+\Db_{I}\Ab'_1 - \Xb_0 -\Db_{I}\Ab'_2}_F^2 \\
    &=& \norm{\Db_I\Ab'_1-\Db_I\Ab'_2}_F^2 \ .
  \end{eqnarray}

The constructions of $\Ab'$ in (\ref{eqn:poiss:classform}), and the class $\bar\cA$ in (\ref{eqn:A:poisson}) imply that the number of degrees of freedom in the sparse matrix $\Ab'$ (which can take values 0 or $a_0$) is restricted to $(k-n_2)$. The Varshamov-Gilbert bound (cf. Lemma 2.9 in \cite{tsybakov2008introduction}) can be easily applied to the set of matrices of the form $\tilde{\Xb}=\Db_I\Ab'$ for $\Ab\in\bar\cA$, and this coupled with (\ref{eqn:poiss:frobeniusDist}) guarantees the existence of a subset $\cX^0_A\subseteq\cX_A$ with cardinality Card($\cX^0_A$) $\geq2^{(k-n_2)/8}+1$, containing the $n_1\times n_2$ reference matrix $\Xb_0=\Xmin\mathbf{1}_{n_1}\cdot\mathbf{1}_{n_2}^T$, such that for any two distinct elements $\Xb_1,\Xb_2\in\cX^0_A$ we have,
\begin{eqnarray}\label{eqn:poiss:condn1_A}
\nonumber 
\norm{\Xb_1-\Xb_2}_F^2 &\geq&
 \left(\frac{k-n_2}{8}\right)\left\lfloor\frac{n_1}{r'}\right\rfloor a_0^2\\
\nonumber
  &=&\left(\frac{k-n_2}{8}\right)\Bigg\lfloor\frac{n_1}{\big\lceil\frac{k-n_2}{n_2}\big\rceil}\Bigg\rfloor \min\left\{\mathrm{A^2_{max}},\gamma_a^2\:\frac{\Xmin}{\Delta(k-n_2,n_2)}\left(\frac{k-n_2}{m}\right)\right\}\\
\nonumber
  &\geq& \left(\frac{n_1n_2}{32}\right)\underbrace{\left(\frac{(k-n_2)\wedge n_2}{n_2}\right)}_{=\Delta(k-n_2,n_2)} \min\left\{\mathrm{A^2_{max}},\gamma_a^2\:\frac{\Xmin}{\Delta(k-n_2,n_2)}\left(\frac{k-n_2}{m}\right)\right\}\\
  &=& \left(\frac{n_1n_2}{32}\right) \min\left\{\Delta(k-n_2,n_2)\mathrm{A^2_{max}},\gamma_a^2\:\Xmin\left(\frac{k-n_2}{m}\right)\right\},
  \end{eqnarray}
  where the second to last inequality comes from the fact that $\lfloor x\rfloor \geq x/2$ when $x\geq1$.

The joint pmf of the set of $|\cS|$ observations (conditioned on the true underlying matrix) can be conveniently written as a product of Poisson pmfs using the independence criterion as,
 \begin{equation}\label{eqn:Poisson_pmf}
   p_{\Xb^*_{\cS}}(\Yb_{\cS})= \prod_{(i,j)\in\cS}\frac{(\Xb^*_{i,j})^{Y_{i,j}}e^{-\Xb^*_{i,j}}}{(Y_{i,j})!}.
 \end{equation}

 For any $\Xb\in\cX^0_A$, the KL divergence of $\PP_{\Xb_0}$ from $\PP_\Xb$ where $\Xb_0$ is the reference matrix (whose entries are all equal to $\Xmin$) is obtained by using an intermediate result from \cite{raginsky2010compressed} giving
 \begin{eqnarray}
 \nonumber
  K(\PP_\Xb,\PP_{\Xb_0}) 
  &=& \EE_{\cS}\left[\sum_{i,j} K(\PP_{\Xb_{i,j}},\PP_{\Xb_{0_{i,j}}}) \right] \\
  \nonumber
  &=& \frac{m}{n_1n_2}\sum_{i,j} \left\{\Xb_{i,j}\log\left(\frac{\Xb_{i,j}}{\Xmin}\right) -\Xb_{i,j}+\Xmin \right\}.
\end{eqnarray}
 
Using the inequality $\log t\leq (t-1)$, we can bound the KL divergence as
\begin{eqnarray}\label{eqn:KLdiv:Poisson1}
 \nonumber
  K(\PP_\Xb,\PP_{\Xb_0}) &\leq& \frac{m}{n_1n_2} \sum_{i,j} \left\{\Xb_{i,j}\left(\frac{\Xb_{i,j}-\Xmin}{\Xmin}\right)-\Xb_{i,j}+\Xmin \right\}\\
  &=& \frac{m}{n_1n_2} \sum_{i,j} \frac{(\Xb_{i,j}-\Xmin)^2}{\Xmin} 
\end{eqnarray}
To further upper bound the RHS of (\ref{eqn:KLdiv:Poisson1}), we note that the number of entries greater than $\Xmin$ in any $\Xb\in\cX^0_A$ is at most $n_1(n_2\wedge(k-n_2))$ by the construction of sets $\cX_A$ and $\bar{\cA}$ in (\ref{def:poiss:calX_A}),(\ref{eqn:A:poisson}) respectively. Hence we have
\begin{eqnarray}
\label{eqn:KLdiv:Poisson2}
	K(\PP_\Xb,\PP_{\Xb_0}) &\leq& m\ \frac{(a_0+\Xmin-\Xmin)^2}{\Xmin} \underbrace{\left(\frac{n_2\wedge(k-n_2)}{n_2}\right)}_{=\Delta(k-n_2,n_2)} \\
\label{eqn:KLdiv:Poisson}
	&=& m\ \frac{\Delta(k-n_2,n_2) a^2_0}{\Xmin},
\end{eqnarray}
where (\ref{eqn:KLdiv:Poisson2}) uses the fact that by construction, entries of the matrices in $\cX_A^0$ are upper bounded by $a_0+\Xmin$. From (\ref{eqn:KLdiv:Poisson}) we can see that 
\begin{equation}\label{eqn:condn2_A:Poisson}
  \frac{1}{\text{Card}(\cX^{0}_{A}) -1}\sum_{\Xb\in\cX^0_A} K(\PP_\Xb,\PP_{\Xb'}) \leq\alpha\log(\text{Card}(\cX^0_A)-1),
\end{equation}
is satisfied for any $0<\alpha<1/8$ by choosing $\gamma_a<\frac{\sqrt{\alpha\log 2}}{2\sqrt{2}} $. Equations (\ref{eqn:poiss:condn1_A}) and (\ref{eqn:condn2_A:Poisson}) imply we can apply Theorem \ref{thm:tsybakov} (where the Frobenius error has been as used the semi-distance function) to yield
\begin{equation}\label{eqn:result_A:Poisson}
  \inf\limits_{\hat{\Xb}}\sup\limits_{\Xb^*\in\cX_A} \PP_{\Xb^*}\left(\frac{\norm{\hat\Xb-\Xb^*}_F^2}{n_1n_2}\geq \frac{1}{64}\cdot \min\left\{\Delta(k-n_2,n_2)\mathrm{A^2_{max}},\gamma_a^2\:\Xmin\left(\frac{k-n_2}{m}\right)\right\}
  \right)\geq \beta
\end{equation}
for some absolute constant $\beta\in(0,1)$.

%--------------------------------------------------------------------------------------
We use arguments similar to the previous case to establish lower bounds using the dictionary term $\Db$. Again, the key factor in the construction of a packing set is to ensure that  entries of the matrices are bounded away from zero (and in fact $\geq\Xmin$). For this let us first define the finite class, $\cX_D\subseteq\cX'(r,k,\Amax,\Xmin)$ as
\begin{equation}\label{defn:calX_D:Poisson}
\cX_{D}\triangleq \left\{\Xb=(\Db_{\delta}+\Db)\Ab:\: \Db\in\bar\cD, \, \Ab=\Amax\left(\begin{array}{c|c|c|c}
\Ib_r&\cdots&\Ib_r&\Psi_D
\end{array}\right) \in\cA, \right\},
\end{equation}
where, $\Ib_r$ denotes the $r\times r$ identity matrix, $\Psi_D$ is an $r\times \left(n_2-r\lfloor n_2/r\rfloor\right)$ matrix given by $\Psi_D=\left(\begin{array}{c}
\Ib_D \\ \hline
\mathbf{0}_D
\end{array}\right)$
and, $\Ib_D$ is the identity matrix of dimension $\left(n_2-r\lfloor n_2/r\rfloor\right)$ and $\mathbf{0}_D$ is the $\left(r-n_2+ r\lfloor n_2/r\rfloor\right)\times \left(n_2-r\lfloor n_2/r\rfloor\right)$ zero matrix, $\bar\cD\subseteq\cD$ and $\Db_\delta\in\cD$ are defined as
\begin{eqnarray}
\label{eqn:defn_calDbar:Poisson}
\nonumber
	\bar\cD &\triangleq& \left\{\Db\in\RR^{n_1\times r}:\, \Db_{i,j}\in\left\{0,d_0\right\},\, \forall (i,j)\in[n_1]\times [r]\right\},\ \& \\
\nonumber
	(\Db_{\delta})_{i,j} &\triangleq& \frac{\Xmin}{\Amax} = \delta \quad \forall (i,j),
\end{eqnarray}
The above definition of $\Ab$ (with the block identity matrices and $\Psi_D$) ensures that entries of the members in our packing set $\cX_D$, are greater than or equal to $\Xmin$. We can see that for any $\Xb\in\cX_D$ we have
\begin{eqnarray}
\nonumber
	\Xb &=& \Db_\delta\Ab + \Db\Ab, \\
\nonumber
\label{eqn:poiss:classformD}
		&=& \underbrace{\delta\Amax\mathbf{1}_{n_1}\cdot\mathbf{1}_{n_2}^T}_{=\Xb_0} +  \Db\Ab.
\end{eqnarray}

Thus we can appeal to the Varshamov-Gilbert bound (cf. Lemma 2.9 in \cite{tsybakov2008introduction}) for matrices of the form $\tilde{\Xb}=\Db\Ab$, where $\Db\in\bar\cD$ and $\Ab$ is defined in (\ref{defn:calX_D:Poisson}), to guarantee the existence of a subset $\cX_D^0\subseteq\cX_D$ with cardinality Card($\cX^0_D$) $\geq2^{n_1r/8}+1$, containing the $n_1\times n_2$ reference matrix $\Xb_0=\Xmin\mathbf{1}_{n_1}\cdot\mathbf{1}_{n_2}^T$, such that for any two distinct elements $\Xb_1,\Xb_2\in\cX^0_D$ we have,
\begin{eqnarray}\label{eqn:poiss:condn1_D}
\nonumber
	\norm{\Xb_1-\Xb_2}_F^2 &\geq&
	\left(\frac{n_1r}{8}\right)\left\lfloor\frac{n_2}{r}\right\rfloor\mathrm{A^2_{max}} d_0^2 \\
	&\geq& \left(\frac{n_1n_2}{16}\right) \min\left\{(1-\delta)^2\mathrm{A^2_{max}},\gamma^2_d\Xmin\left(\frac{n_1r}{m}\right)\right\}.
\end{eqnarray}

For any $\Xb\in\cX^0_D$, the KL divergence of $\PP_{\Xb_0}$ from $\PP_\Xb$ where $\Xb_0$ is the reference matrix (whose entries are all equal to $\Xmin$) can be upper bounded using  (\ref{eqn:KLdiv:Poisson1}) by
\begin{eqnarray}\label{eqn:KLdiv:Poisson_D1}
\nonumber
K(\PP_\Xb,\PP_{\Xb_0}) &\leq& \frac{m}{n_1n_2} \sum_{i,j} \frac{(\Xb_{i,j}-\Xmin)^2}{\Xmin} \\
&\leq& m\: \frac{d_0^2\:\mathrm{A^2_{max}}}{\Xmin},
\end{eqnarray}
where (\ref{eqn:KLdiv:Poisson_D1}) uses the fact that by construction, entries of the matrices in $\cX^0_D$ are upper bounded by $d_0+\Xmin$. From (\ref{eqn:KLdiv:Poisson_D1}) we can see that 
\begin{equation}\label{eqn:condn2_D:Poisson}
\frac{1}{\text{Card}(\cX^{0}_{D}) -1}\sum_{\Xb\in\cX^0_D} K(\PP_\Xb,\PP_{\Xb'}) \leq\alpha'\log(\text{Card}(\cX^0_D)-1),
\end{equation}
is satisfied for any $0<\alpha'<1/8$ by choosing $\gamma_d<\frac{\sqrt{\alpha'\log 2}}{2\sqrt{2}} $. Equations (\ref{eqn:poiss:condn1_D}) and (\ref{eqn:condn2_D:Poisson}) imply we can apply Theorem \ref{thm:tsybakov} (where the Frobenius error has been as used the semi-distance function) to yield
\begin{equation}\label{eqn:result_D:Poisson}
\inf\limits_{\hat{\Xb}}\sup\limits_{\Xb^*\in\cX_A} \PP_{\Xb^*}\left(\frac{\norm{\hat\Xb-\Xb^*}_F^2}{n_1n_2}\geq \frac{1}{64}\cdot \min\left\{(1-\delta)^2\mathrm{A^2_{max}},\gamma_d^2\:\Xmin\left(\frac{n_1r}{m}\right)\right\}
\right)\geq \beta',
\end{equation}
for some absolute constant $\beta'\in(0,1)$. Using (\ref{eqn:result_A:Poisson}) and (\ref{eqn:result_D:Poisson}), and by applying Markov's inequality we directly get the result presented in Theorem \ref{thm:poisson}, thus completing the proof.  \hfill \qedsymbol
%======================================================================================================

\bibliographystyle{IEEEbib}
\bibliography{minimax_lb_ref}

\begin{thebibliography}{10}

\bibitem{candes2009exact}
E.~J. Cand{\`e}s and B.~Recht,
\newblock ``Exact matrix completion via convex optimization,''
\newblock {\em Foundations of Computational mathematics}, vol. 9, no. 6, pp.
  717--772, 2009.

\bibitem{candes2010power}
E.~J. Cand{\`e}s and T.~Tao,
\newblock ``The power of convex relaxation: Near-optimal matrix completion,''
\newblock {\em IEEE Trans. Information Theory}, vol. 56, no. 5, pp. 2053--2080,
  2010.

\bibitem{keshavan2010matrix}
R.~Keshavan, A.~Montanari, and S.~Oh,
\newblock ``Matrix completion from a few entries,''
\newblock {\em IEEE Trans. Information Theory}, vol. 56, no. 6, pp. 2980--2998,
  2010.

\bibitem{recht2011simpler}
B.~Recht,
\newblock ``A simpler approach to matrix completion,''
\newblock {\em The Journal of Machine Learning Research}, vol. 12, pp.
  3413--3430, 2011.

\bibitem{gross2011recovering}
D.~Gross,
\newblock ``Recovering low-rank matrices from few coefficients in any basis,''
\newblock {\em IEEE Trans. Information Theory}, vol. 57, no. 3, pp. 1548--1566,
  2011.

\bibitem{keshavan2009matrix}
R.~Keshavan, A.~Montanari, and S.~Oh,
\newblock ``Matrix completion from noisy entries,''
\newblock in {\em Advances in Neural Information Processing Systems}, 2009, pp.
  952--960.

\bibitem{candes2010matrix}
E.~J. Cand{\`e}s and Y.~Plan,
\newblock ``Matrix completion with noise,''
\newblock {\em Proceedings of the IEEE}, vol. 98, no. 6, pp. 925--936, 2010.

\bibitem{koltchinskii2011nuclear}
V.~Koltchinskii, K.~Lounici, and A.~B. Tsybakov,
\newblock ``Nuclear-norm penalization and optimal rates for noisy low-rank
  matrix completion,''
\newblock {\em The Annals of Statistics}, vol. 39, no. 5, pp. 2302--2329, 2011.

\bibitem{rohde2011estimation}
A.~Rohde and A.~B. Tsybakov,
\newblock ``Estimation of high-dimensional low-rank matrices,''
\newblock {\em The Annals of Statistics}, vol. 39, no. 2, pp. 887--930, 2011.

\bibitem{cai2013matrix}
T.~T. Cai and W.~Zhou,
\newblock ``Matrix completion via max-norm constrained optimization,''
\newblock {\em Electronic Journal of Statistics}, vol. 10, no. 1, pp.
  1493--1525, 2016.

\bibitem{klopp2014noisy}
O.~Klopp,
\newblock ``Noisy low-rank matrix completion with general sampling
  distribution,''
\newblock {\em Bernoulli}, vol. 20, no. 1, pp. 282--303, 2014.

\bibitem{lafond2015low}
J.~Lafond,
\newblock ``Low rank matrix completion with exponential family noise.,''
\newblock in {\em COLT}, 2015, pp. 1224--1243.

\bibitem{davenport20141}
M.~A. Davenport, Y.~Plan, E.~van~den Berg, and M.~Wootters,
\newblock ``1-bit matrix completion,''
\newblock {\em Information and Inference}, vol. 3, no. 3, pp. 189--223, 2014.

\bibitem{plan2014high}
Y.~Plan, R.~Vershynin, and E.~Yudovina,
\newblock ``High-dimensional estimation with geometric constraints,''
\newblock {\em Information and Inference}, p. iaw015, 2016.

\bibitem{soni2014estimation}
A.~Soni and J.~Haupt,
\newblock ``Estimation error guarantees for {P}oisson denoising with sparse and
  structured dictionary models,''
\newblock in {\em IEEE Intl Symposium on Information Theory}. IEEE, 2014, pp.
  2002--2006.

\bibitem{candes2011robust}
E.~J. Cand{\`e}s, X.~Li, Y.~Ma, and J.~Wright,
\newblock ``Robust principal component analysis?,''
\newblock {\em Journal of the ACM (JACM)}, vol. 58, no. 3, pp. 11, 2011.

\bibitem{hsu2011robust}
D.~Hsu, S.~M. Kakade, and T.~Zhang,
\newblock ``Robust matrix decomposition with sparse corruptions,''
\newblock {\em IEEE Trans. Information Theory}, vol. 57, no. 11, pp.
  7221--7234, 2011.

\bibitem{xu2012robust}
H.~Xu, C.~Caramanis, and S.~Sanghavi,
\newblock ``Robust {PCA} via outlier pursuit,''
\newblock {\em IEEE Trans. Information Theory}, vol. 58, no. 5, pp.
  3047–--3064, 2012.

\bibitem{chen2013low}
Y.~Chen, A.~Jalali, S.~Sanghavi, and C.~Caramanis,
\newblock ``Low-rank matrix recovery from errors and erasures,''
\newblock {\em IEEE Trans. Information Theory}, vol. 59, no. 7, pp. 4324--4337,
  2013.

\bibitem{soni2014noisy}
A.~Soni, S.~Jain, J.~Haupt, and S.~Gonella,
\newblock ``Noisy matrix completion under sparse factor models,''
\newblock {\em IEEE Trans. Information Theory}, vol. 62, no. 6, pp. 3636--3661,
  2016.

\bibitem{tsybakov2008introduction}
A.~B. Tsybakov,
\newblock {\em Introduction to nonparametric estimation},
\newblock Springer, 2008.

\bibitem{klopp2014robust}
O.~Klopp, K.~Lounici, and A.~B. Tsybakov,
\newblock ``Robust matrix completion,''
\newblock {\em Probability Theory and Related Fields}, pp. 1--42, 2014.

\bibitem{haupt2014sparse}
J.~Haupt, N.~Sidiropoulos, and G.~Giannakis,
\newblock ``Sparse dictionary learning from 1-bit data,''
\newblock in {\em IEEE Intl. Conference on Acoustics, Speech and Signal
  Processing}. IEEE, 2014, pp. 7664--7668.

\bibitem{ribeiro2006bandwidth}
A.~Ribeiro and G.~B. Giannakis,
\newblock ``Bandwidth-constrained distributed estimation for wireless sensor
  networks-part i: Gaussian case,''
\newblock {\em IEEE Trans. Signal Processing}, vol. 54, no. 3, pp. 1131--1143,
  2006.

\bibitem{luo2005universal}
Z.~Q. Luo,
\newblock ``Universal decentralized estimation in a bandwidth constrained
  sensor network,''
\newblock {\em IEEE Trans. Information Theory}, vol. 51, no. 6, pp. 2210--2219,
  2005.

\bibitem{lafond2014probabilistic}
J.~Lafond, O.~Klopp, E.~Moulines, and J.~Salmon,
\newblock ``Probabilistic low-rank matrix completion on finite alphabets,''
\newblock in {\em Advances in Neural Information Processing Systems}, 2014, pp.
  1727--1735.

\bibitem{raginsky2010compressed}
M.~Raginsky, R.~M. Willett, Z.~T. Harmany, and R.~F. Marcia,
\newblock ``Compressed sensing performance bounds under {P}oisson noise,''
\newblock {\em IEEE Trans. Signal Processing}, vol. 58, no. 8, pp. 3990--4002,
  2010.

\bibitem{kolaczyk2004multiscale}
E.~D. Kolaczyk and R.~D. Nowak,
\newblock ``Multiscale likelihood analysis and complexity penalized
  estimation,''
\newblock {\em Annals of Statistics}, pp. 500--527, 2004.

\bibitem{jiang2014minimax}
X.~Jiang, G.~Raskutti, and R.~Willett,
\newblock ``Minimax optimal rates for {P}oisson inverse problems with physical
  constraints,''
\newblock {\em IEEE Trans. Information Theory}, vol. 61, no. 8, pp. 4458--4474,
  2015.

\bibitem{kolda2009tensor}
T.~G. Kolda and B.~W. Bader,
\newblock ``Tensor decompositions and applications,''
\newblock {\em SIAM review}, vol. 51, no. 3, pp. 455--500, 2009.

\end{thebibliography}
\end{document}